\pdfoutput=1

\documentclass[11pt]{article}

\usepackage[preprint]{acl}

\usepackage{times}
\usepackage{latexsym}
\usepackage{supertabular}
\usepackage{amsmath}
\usepackage{colortbl}
\usepackage{xcolor}
\usepackage{longtable}
\usepackage{multicol}
\usepackage{mdframed}
\usepackage{float}
\definecolor{lightgray}{gray}{0.9}
\definecolor{lightblue}{rgb}{0.90, 0.92, 0.97}
\definecolor{lightred}{rgb}{0.98,0.89,0.89}
\definecolor{tablebg}{rgb}{0.95,0.95,0.95}
\definecolor{tableline}{rgb}{0.99,0.43,0.40}
\newmdenv[
  backgroundcolor=lightgray,
  linecolor=lightred,
  skipabove=10pt,
  skipbelow=10pt,
  leftmargin=-10pt,
  rightmargin=-30pt,
  innerleftmargin=2pt,
  innerrightmargin=2pt,
  innertopmargin=0pt,
  innerbottommargin=10pt,
  roundcorner=50pt,
]{bluebox}

\usepackage[T1]{fontenc}

\usepackage[utf8]{inputenc}

\usepackage{microtype}

\usepackage{inconsolata}

\usepackage{graphicx}

\usepackage{hyperref}
\title{Can LLMs faithfully generate their layperson-understandable ``self''?: A Case Study in High-Stakes Domains}

\usepackage{authblk}

\author[1]{Arion Das \thanks{These authors contributed equally to this work.}}
\author[2]{Asutosh Mishra \thanks{These authors contributed equally to this work.}}
\author[1]{Amitesh Patel \thanks{These authors contributed equally to this work.}}
\author[3]{Soumilya De \thanks{These authors contributed equally to this work.}}
\author[4]{V. Gurucharan \thanks{These authors contributed equally to this work.}}
\author[3]{Kripabandhu Ghosh \thanks{Corresponding author: \texttt{kripa.ghosh@gmail.com}}}

\affil[1]{Indian Institute of Information Technology, Ranchi, India}
\affil[2]{Indian Institute of Science Education and Research, Berhampur, India }
\affil[3]{Indian Institute of Science Education and Research, Mohanpur, India }
\affil[4]{Collaborative Dynamics, Texas, United States}

\begin{document}
\maketitle
\arrayrulecolor{tableline}
\begin{abstract}
Large Language Models (LLMs) have significantly impacted nearly every domain of human knowledge. However, the explainability of these models esp. to laypersons, which are crucial for instilling trust, have been examined through various skeptical lenses. In this paper, we introduce a novel notion of LLM explainability to laypersons, termed \textit{ReQuesting}, across three high-priority application domains -- law, health and finance, using multiple state-of-the-art LLMs. The proposed notion exhibits faithful generation of explainable layman-understandable algorithms on multiple tasks through high degree of reproducibility. Furthermore, we observe a notable aligment of the explainable algorithms with intrinsic reasoning of the LLMs.
\end{abstract}

\section{Introduction}

\begin{quote}
``{\it I would rather have questions that can't be answered than answers that can't be questioned.}''

\hfill --- Richard Feynman
\end{quote}


In the 1800s Edward Jenner~\cite{Jenner} faced stringent opposition against smallpox vaccination due to multiple reasons primarily stemming from a lack of general understanding of the mechanism of vaccination. Despite the initial opposition, the vaccine was gradually well-accepted across the globe resulting in the 1980 WHO declaration that smallpox was eradicated from earth\footnote{\url{https://www.who.int/health-topics/smallpox}}. There are many such watershed moments in history, where a novel phenomenon faced opposition branching out of its lack of understanding to the common people e.g. {\it Heliocentric model and Galileo}~\cite{Galileo}, {\it Industrial Revolution, Luddite movement}~\cite{IndustrRev} etc. Such skepticism, due to a lack of general understanding, is more pronounced when the area in question is of high stake e.g. health, law, etc.

Large Language Models (LLMs) have shown promise in performing numerous tasks, encompassing both computational and semantic processing. As a result, they have found applications in diverse domains, from customer service and content creation to high-stakes domains like law \cite{vats2023llms}, healthcare \cite{thirunavukarasu2023} and finance \cite{zhang2023enhancing} (more details in Appendix \ref{app:literature}). Nevertheless, {\it explainability is especially critical to ensure trust on LLM deployment to these high-stakes domains}~\cite{compas, healthtrustLLM}. However, as LLMs are sophisticated AI models, which are highly intractable due to their complex architecture, the trustworthiness of LLMs is the foremost concern, as discussed in \cite{sarker2024llm} and examined in \cite{huang2023trustgpt}. A standard approach towards trustworthy AI, is to improve the explainability \cite{10.1145/3639372} and interpretability \cite{templeton2024scaling} of LLM. Another popular approach as exemplified by \cite{band2024linguistic}, illustrates that reducing hallucinations can lead to a more trustworthy AI. Many attempts have been made to quantify the uncertainty \cite{yadkori2024believe} and confidence \cite{xiong2024llms} associated with the language model output. However, to our knowledge, {\it there exists no work that engage in generation of ``layperson--understandable'' algorithmic representation of the working mechanism from the LLMs}. On this note, we start with a novel research question {\bf RQ1: Can the LLMs generate a faithful layperson-understandable algorithmic representation of the working mechanism?}

\begin{figure}[ht] 
  \centering    
\includegraphics[width=5cm, height=5cm]{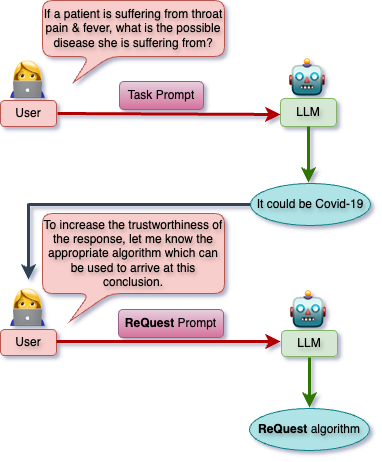} 
  \caption{A schematic representing the ReQuesting technique.} 
    \label{fig:ReQuesting}
\end{figure}

To this end, this paper introduces a novel approach termed `ReQuesting', representing the act of \textbf{Re}peated \textbf{Quest}ioning through prompt requests to the LLM (detailed and defined more formally in Section \ref{sec:ReQuest}). 
Figure \ref{fig:ReQuesting} represents the ReQuesting technique employed in this paper. 

ReQuest aims to generate a \textbf{higher-order abstracted conceptual model} (a.k.a. a common person understandable ``algorithm'') faithfully representing the underlying working that LLMs use for complex tasks like statute prediction. While this does not imply the LLM explicitly uses this abstract algorithm, the aim is to provide a {\it faithful} mechanism across tasks, as evidenced by similar outputs from ReQuest and direct task prompts. This higher-order abstraction is supposedly more explainable to laypersons and is therefore useful for professionals in fields like law or medicine. Our novel approach uses natural language prompting to create this abstraction, making it accessible to domain experts and less computationally intensive than current methods, which are often limited to open-source LLMs \cite{10.1145/3639372}. Although LLM self-explanations do not aim to reveal the underlying mathematical algorithms, it is worth noting that \cite{huang2023largelanguagemodelsexplain} demonstrates that ChatGPT's self-explanations are on par with traditional explanation methods such as occlusion or LIME saliency maps~\cite{lime2016}, making them reasonably reliable, which lends further support to our approach. The key difference between Chain-of-Thought (CoT) prompting and ReQuesting lies in their approach to task execution. CoT encourages the model to perform step-by-step reasoning, thereby generating problem-solving steps from a human perspective \cite{wei2023chainofthought}, which often mimics human approaches and requires human annotation and evaluation. In contrast, ReQuesting prompts the model to automatically identify and employ the most appropriate algorithm for the task {\it without requiring human-evaluation}.

To dig deep deeper into `ReQuesting', we ask another research question {\bf RQ2: How to determine ``faithfulness''?} In this paper, we have utilized ReQuesting to benchmark the response faithfulness through reproducibility of state-of-the-art open-source LLMs like \texttt{gemini}, \texttt{llama3} etc. and computed reproducibility through measures (defined in Section \ref{sec:ReQuest}) based on Macro F1 and Jaccard score \cite{Jaccard} on three high-stake domains: law (Section \ref{sec:legal}), finance (Section \ref{sec:finance}) and health (Section \ref{sec:health}). {\it Table \ref{fig:PerRR} shows consistently high reproducibility and hence reasonable faithfulness across the three domains, confirming the efficacy of the ReQuest setup}.

Finally, we raise the research question: {\bf RQ3: Does the ReQuest algorithm align with the ``intrinsic reasoning'' of the LLM?}
Taking the cue from \cite{CoTRWOP} we explore alternative paths of generation to investigate potential internal reasoning of LLMs during the task and the underlying alignment with the `ReQuest' prompt (discussed in Section \ref{sec:intrinsic}).




\section{RQ1: Can the LLMs generate a faithful layperson-understandable algorithmic representation of the working mechanism?}

To answer this question, we propose a novel evaluation setup that is designed to interact with the LLMs to elicit a representative algorithm that can faithfully represent the underlying working mechanism. This brings us to our next research question: {\bf RQ2: How to determine ``faithfulness''?} In this course, we attempt to realize faithfulness through reproducibility. 

\subsection{ReQuest: Faithfulness through Reproducibility}\label{sec:ReQuest}

In this section, we introduce a novel {\it regime of prompts}: a new perspective on generative AI reproducibility. The proposed regime is composed of the following three types of prompts:

     \noindent 1. \textbf{Task Prompt:} This is to prompt an LLM (e.g. \texttt{gemini}) to perform a Machine Learning task (e.g. legal statute prediction), say $\mathcal{T}$.\\
     \noindent 2. {\bf ReQuest Prompt:} This is a follow-up prompt after the {\it Task Prompt} has produced an output in response to the task. This  {\it ReQuest Prompt ($\mathcal{R}$)} prompts the LLM to generate an algorithm ($\mathcal{A}$) (refer Table \ref{tab:request_sample}) that can perform the aforesaid task $\mathcal{T}$. 

\begin{table}[]
    \centering
    \small
    \begin{tabular}{|p{7.7cm}|}
    \hline
        1. **Read and understand the fact statement.** Identify the key legal issues in the fact statement.\\
        2. **Identify the relevant area of law.** For example, if the fact statement involves a criminal offense, then the relevant area of law is criminal law.\\
        3. **Research the relevant statutes.** This may involve using a legal database or consulting with a legal professional.\\
        4. **Analyze the statutes and determine which ones are applicable to the fact statement.\\
    
    \hline
    \end{tabular}
    \caption{Sample ReQuest Algorithm for Statute Prediction. For complete ReQuest algorithm refer Table \ref{tab:statute_algo_gemini}. }
    \label{tab:request_sample}
\end{table}
    
    Please note that the authors are aware of the fact that LLMs are complex Deep Learning models that leverage internal architecture to generate a solution to the task. However, given the unprecedented generative capabilities of such models \cite{codeNeurips2023}, we examine whether the generative models can enlighten the user by producing the reasoning in the form of an algorithm that can be deployed to perform the task $\mathcal{T}$. 
    
    The idea is, that the more reproducible this algorithm ($\mathcal{A}$), the more trustworthy the LLM is. In other words, to a layperson (who can be a legal practitioner, a doctor, or a stock market analyst), we will be in a better position to explain the output of the LLM through the algorithm.\\
     \noindent 3. {\bf Robustness Check Prompt:} We further prompt the LLM to execute the same task $\mathcal{T}$ but by employing the algorithm $\mathcal{A}$. Note that \cite{zheng2024executing} have also tried to run algorithms written in natural language and python codes on LLMs, but unlike us, they did not check reproducibility and that too on LLM generated codes.  We evaluate the reproducibility through some evaluation measures (discussed shortly) not only on the same LLM (that produced $\mathcal{A}$) -- which we refer to as {\it intra-LLM} setup, but also in another LLM -- which we refer to as {\it inter-LLM} setup. 

Here we proceed through the prevalent reasoning hypothesis: {\it Two empirical processes are comparable if they produce comparable results on the identical experimental setup}. That is, if the {\bf Robustness Check Prompt} produces comparable results with the original LLM prompt ({\bf Task Prompt}), then $\mathcal{A}$ may be considered as a ``faithful'' (and explainable) representation of the complex mechanism performed through the \textbf{Task Prompt}.


\begin{figure}[ht]
  \centering
  \includegraphics[width=1\columnwidth]{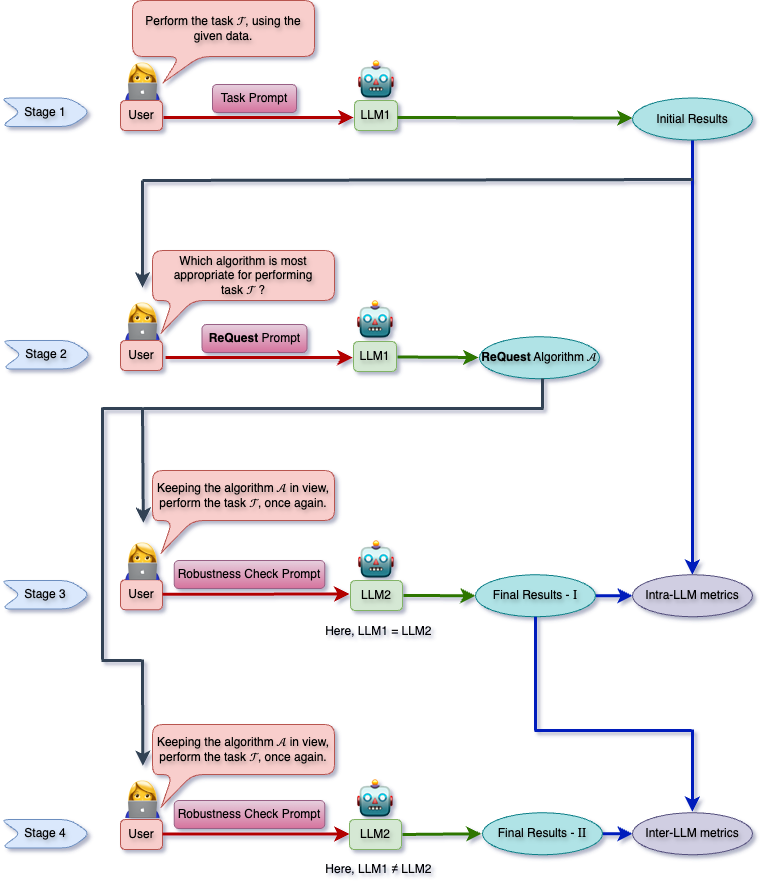} 
  \caption{A detailed schematic depicting the workflow utilized in this study.}
  \label{fig:workflow}
\end{figure}


The overall workflow of the `ReQuest' regime is shown in Figure \ref{fig:workflow}. An example for the health domain is shown in Figure \ref{fig:request_eg} (in Appendix). We will discuss the detailed prompts in the proposed regime in the following sections.

\subsubsection{Evaluation measures}
\label{sec:rep}

To evaluate the reproducibility of LLMs on task $\mathcal{T}$, we propose the following evaluation measures. {
We are not aware of any other suitable metric for our task. Further discussion on the state of the art reproducibility metric has been provided in Appendix \ref{app:reproducibility}, which shows the need to propose new metrics to evaluate our novel task.}

     \noindent 1. {\bf Performance Reproduction ratio (PerRR)}: This is the percentage F1-score reproduced on task $\mathcal{T}$. We further define the variants of PerRR as follows:
    \begin{itemize}
        \item PerRR\_LLM1P\_LLM2: Percentage score (Macro F1) reproducibility between the prompt of LLM1 (e.g. \texttt{gemini}) and the {\it ReQuest} algorithm $\mathcal{A}$ produced by LLM1 on LLM2 (LLM2 could be same as LLM1 or could be different as well; if they are the same, it is the Intra-LLM setup, otherwise it is the Inter-LLM setup)
        \item PerRR\_LLM1A\_LLM2: Percentage score (Macro F1) difference between the {\it ReQuest} algorithm $\mathcal{A}$ produced by LLM1 (e.g. \texttt{gemini}) and that when the same algorithm $\mathcal{A}$ is invoked on LLM2.
    \end{itemize}
    
    \vspace{10mm}
    The {\bf PerRR} variants are defined as follows:
    \vspace{-3mm}

    \begin{equation*}
    PerRR\_l_{1}\_l_{2} = 100 - \frac{|l\_1^{macf1}-l\_2^{macf1}|}{l\_1^{macf1}}*100
    \end{equation*}
    where, $l_1$ is LLM1P or LLM1A, $l_2$ is LLM2; $macf1$ is Macro F1. 
    
    The aim of {\bf PerRR} is to measure the overall discrepancy between the prediction performance. However, this may not necessarily reveal the finer-level discrepancies in predictions esp. at the test data level. To capture the latter, we define the following measures.\\
     \noindent 2. {\bf Prediction reproduction ratio (PreRR)}: This is designed to measure the Jaccard overlap~\cite{Jaccard} of exact predictions for each data point.

    \begin{itemize}
        \item PreRR\_LLM1P\_LLM2: This is Jaccard overlap between the predicted classes by the prompt of LLM1 and the {\it ReQuest} algorithm $\mathcal{A}$, produced by LLM1 on LLM2, averaged over all test data points.
        \item PreRR\_LLM1A\_LLM2: This is Jaccard overlap between the predicted classes by the {\it ReQuest} algorithm $\mathcal{A}$ produced by LLM1 and that when the same algorithm $\mathcal{A}$ is `executed' on LLM2, averaged over all test data points.
    \end{itemize}
    The {\bf PreRR} variants are defined as follows:
    \vspace{-2mm}
    
    \begin{equation*}
    \centering
       PreRR\_l_{1}\_l_{2} = \frac{\sum_{d \in D} Jaccard(pc(l_1^d), pc(l_2^d))}{|D|}
    \end{equation*}

    where, $pc(l_1^d)$ is the predicted classes by the LLM $l_1$ for the data point $d$ $\in$ $D$ (set of all test data points); $l_1$ is LLM1P or LLM1A, $l_2$ is LLM2.
    
    As an example, if the classes predicted by an LLM1 prompt are \{$c_1$, $c_2$\} and those predicted by the corresponding ReQuest Algorithm $\mathcal{A}$ are \{$c_1$, $c_2$\, $c_3$\}, {\bf PreRR} score will be the Jaccard overlap of these two sets, which will be $\frac{2}{3}$ i.e. 0.667 (approx). Note that this measure is {\it independent} of the gold standard (as that is already measured by Macro F1) and measures the deviations between the predictions (please see Section \ref{sec:PreRRvsPerRR} for more details).

The above evaluation measures would be calculated within each LLM -- \texttt{gemini-1.0-pro} (via Google's makersuite API) and \texttt{llama3-70b} (via Groq API) and across.

\begin{figure*}[ht] 
  \centering    
\includegraphics[width=15cm, height=6.5cm]{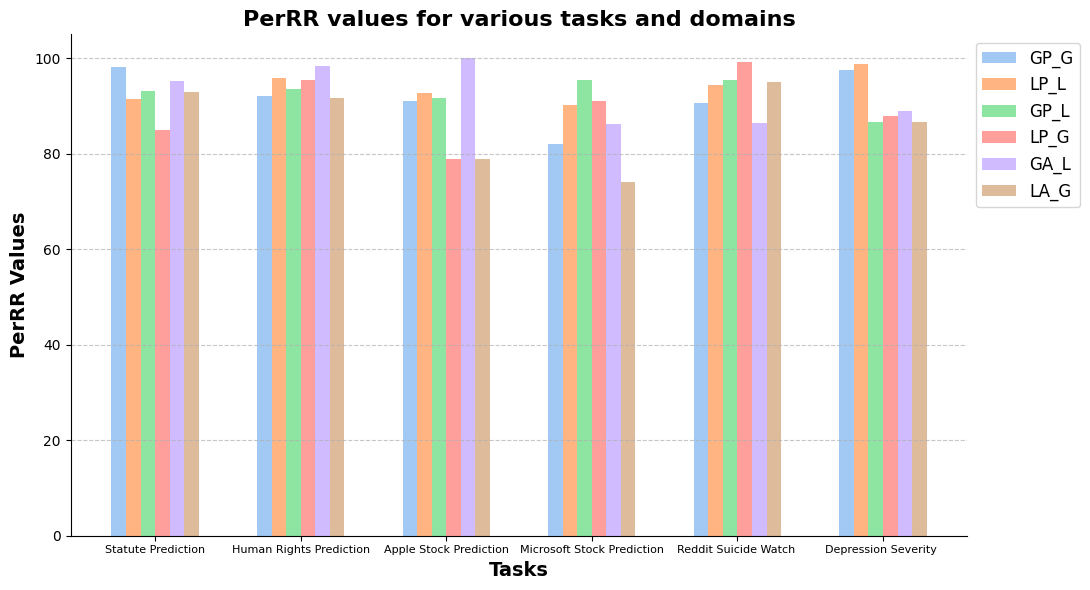} 
  \caption{PerRR values across tasks and LLMs. The consistently high values demonstrate high reproducibility and hence high degree of faithfulness of the ReQuest paradigm.}
    \label{fig:PerRR}
\end{figure*}

\begin{table*}[h]
\centering
\makebox[\textwidth][c]{ 
\rowcolors{1}{tablebg}{white}
\scriptsize
\begin{tabular}{|c|c|c|c|c|c|c|c|c|c|}
\hline
\rowcolor{lightred}
\multicolumn{10}{|c|}{Statute Prediction~\citep{vats2023llms} } \\ 
\hline
\multicolumn{5}{|c|}{Intra-LLM} & \multicolumn{5}{|c|}{Inter-LLM} \\ 
\hline
\multicolumn{3}{|c|}{Macro F1} &  Percentage & Ratio & Macro F1 &  Percentage & Ratio &  Percentage & Ratio\\ 
\hline
{\bf baseline (GPT-3.5)} & {\bf Gemini} & {\bf ReQuest algo} &  {\bf PerRR\_GP\_G} & {\bf PreRR\_GP\_G} & {\bf Gemini $\rightarrow$ Llama} & {\bf PerRR\_GP\_L} & {\bf PreRR\_GP\_L} & {\bf PerRR\_GA\_L} & {\bf PreRR\_GA\_L}\\
\hline
0.2586            & 0.2912         &  0.2967        & 98.111 {\color{green}$\uparrow$} & 0.5188 & 0.2912 & 93.166 {\color{green}$\uparrow$} &  0.4487 & 95.146 {\color{green}$\uparrow$} & 0.3891\\
\hline                   
{\bf baseline (GPT-3.5)} & {\bf Llama} & {\bf ReQuest algo} &  {\bf PerRR\_LP\_L} & {\bf PreRR\_LP\_L} & {\bf Llama $\rightarrow$ Gemini} & {\bf PerRR\_LP\_G} & {\bf PreRR\_LP\_G} & {\bf PerRR\_LA\_G} & {\bf PreRR\_LA\_G}\\
\hline
0.2586             & 0.3264       & 0.2983         & 91.39  {\color{red}$\downarrow$} & 0.6083 & 0.3264  & 84.988 {\color{red}$\downarrow$} &  0.3880 & 92.993 {\color{red}$\downarrow$} & 0.4103\\
\hline

\hline
\rowcolor{lightred}
\multicolumn{10}{|c|}{Human Rights Prediction~\citep{chalkidis-etal-2019-neural} } \\ 
\hline
\multicolumn{5}{|c|}{Intra-LLM} & \multicolumn{5}{|c|}{Inter-LLM} \\ 
\hline
\multicolumn{3}{|c|}{Macro F1} &  Percentage & Ratio & Macro F1 &  Percentage & Ratio &  Percentage & Ratio\\ 
\hline
{\bf baseline (L-BERT)} & {\bf Gemini} & {\bf ReQuest algo} &  {\bf PerRR\_GP\_G} & {\bf PreRR\_GP\_G} & {\bf Gemini $\rightarrow$ Llama} & {\bf PerRR\_GP\_L} & {\bf PreRR\_GP\_L} & {\bf PerRR\_GA\_L} & {\bf PreRR\_GA\_L}\\
\hline
0.8298           & 0.4231        &  0.3893       & 92.011 {\color{red}$\downarrow$} & 0.9613 & 0.4231 & 93.453 {\color{red}$\downarrow$} &  0.9377 & 98.433 {\color{green}$\uparrow$} & 0.9579\\
\hline                   
{\bf baseline (L-BERT)} & {\bf Llama} & {\bf ReQuest algo} &  {\bf PerRR\_LP\_L} & {\bf PreRR\_LP\_L} & {\bf Llama $\rightarrow$ Gemini} & {\bf PerRR\_LP\_G} & {\bf PreRR\_LP\_G} & {\bf PerRR\_LA\_G} & {\bf PreRR\_LA\_G}\\
\hline
0.8298           & 0.4105       & 0.4275        & 95.859  {\color{green}$\uparrow$} & 0.9609 & 0.4105 & 95.396 {\color{red}$\downarrow$} & 0.9550 & 91.602 {\color{red}$\downarrow$} & 0.9327\\
\hline

\end{tabular}
} 

\caption{Reproducibility: Legal. For Statute Prediction, the results show a decent overall (Macro F1) reproducibility as shown by PerRR, but relatively lower PreRR values, that indicate internal variations in LLMs. However, the Human rights prediction task exhibits high reproducibility performance of the LLMs.}
\label{tab:legal}
\end{table*}

\section{Legal Tasks}\label{sec:legal}


\subsection{Statute Prediction}\label{subsubsec_Statute_dataset}
The Statute Prediction task constitutes the prediction of the pertinent statutes (written law) from a legal situation expressed in textual format. We selected the statute prediction dataset in \cite{vats2023llms} for statute prediction, containing 45 fact texts from cases of Supreme Court of India, with a total of 18 unique statutes to choose from out of which multiple statutes were applicable for a single fact statement. Hence this is a {\it multi-label classification} task. Below is an example excerpt from an input statement and its corresponding statutes:\\
{\bf Input:} ... P4 took out a desi katta and fired at P2 and the bullet hit in the stomach area. ...\\
{\bf Applicable statutes:} Indian Penal Code, 1860\_307, Indian Penal Code, 1860\_302


\subsection{Human Rights Violation Prediction}\label{subsubsec_HR_dataset}
The European Convention of Human Rights\footnote{\url{https://www.echr.coe.int/documents/d/echr/Convention_ENG}} dataset (ECHR\footnote{\url{https://archive.org/details/ECHR-ACL2019}}) \cite{chalkidis-etal-2019-neural} contains approx. 11.5k cases from the ECHR’s public database.\footnote{\url{https://hudoc.echr.coe.int}}. These cases are mapped to those articles and protocols of the European Convention that were violated according to the case description. The dataset also provides a list of facts extracted from the case description using regular expressions (Fig 1 at \cite{chalkidis-etal-2019-neural}). We used these fact statements to predict whether any ECHR article/protocol is violated, making it a {\it binary-classification} task.

Table \ref{tab:legal_egs} (in Appendix) shows examples from these two datasets.

\subsection{Methods}
\noindent {\bf Baselines.} As the task baseline for {\bf statute prediction}, we considered the Macro-F1 score of the best-performing model~(without explanations) reported in \citeauthor{vats2023llms}~(\citeyear{vats2023llms}). We fine-tuned {\it Legal-BERT}\footnote{\url{https://huggingface.co/nlpaueb/legal-bert-base-uncased}} \cite{chalkidis-etal-2020-legal} on the ECHR dataset~\cite{chalkidis-etal-2019-neural} for the {\bf Human Rights Violation} task. 
\\
\noindent {\bf Prompts.} We borrowed the template 2 prompt from the \citeauthor{vats2023llms} as the task prompt for {\bf statute prediction} and introduced minor LLM-specific modifications to ensure consistent output formats. The exact task prompts are provided in Tables \ref{tab:statute_task_prompt_gemini} and \ref{tab:statute_task_prompt_Llama}. The algorithms obtained after ReQuesting the LLMs are listed in Tables \ref{tab:statute_algo_gemini} and \ref{tab:statute_algo_Llama}. The robustness check prompts employing these algorithms are referred to in the Table \ref{tab:robustness_statute_gemini}.

Through the task prompts (Table \ref{tab:HR_task_prompt_gemini}-\ref{tab:HR_task_prompt_Llama}) for the {\bf Human Rights Violation} task, we ensure that the LLMs refer the HR-NET\footnote{\url{http://www.hri.org/docs/ECHR50.html}} website as the source of detailed articles and protocols of ECHR.
Tables \ref{tab:HR_algo_gemini} and \ref{tab:HR_algo_Llama} contain the ReQuest algorithms for \texttt{gemini} and \texttt{llama3} respectively. For robustness check, the LLM is first assigned the role of a ``{\it{bot that strictly follows steps}}'' to provide it the context for following the ReQuest algorithm. The respective robustness check prompts can be found in Tables \ref{tab:robustness_HR_gemini} and \ref{tab:robustness_HR_Llama}. {
Based on the findings of \cite{sclar2024quantifyinglanguagemodelssensitivity} that capitalization influences LLM sensitivity, we capitalized key sections of the prompts to ensure valid and consistent output formats, essential for automatic evaluation of large datasets.}

\subsection{Results}

The results are shown in Table \ref{tab:legal}.

\noindent {\bf Statute Prediction}: 
On the same {\it task prompt} as in \cite{vats2023llms}, we observe that both \texttt{gemini} and \texttt{llama3} outperform the baseline. Next, as we obtain the {\it ReQuest Algorithm} via {\it ReQuest Prompt} like ``What steps did you follow to arrive at these statutes at the end?''. Note that to ensure that the LLM produces a reproducible or deterministic algorithm, we also used sentences like ``Make these steps more deterministic.'' as a part of the {\it ReQuest Prompt}. Once we get an algorithm (which we call the {\it ReQuest Algorithm} $\mathcal{A}$), we prompt an LLM to `run' the same algorithm for the same task (here, statute detection), which we call the {\it Robustness Check Prompt}. Table \ref{tab:statute_algo_gemini} (for \texttt{gemini}, in the Appendix) shows a {\it ReQuest Algorithm} and Table \ref{tab:robustness_statute_gemini} shows {\it Robustness Check Prompt} (in the Appendix).
\noindent {\bf Human Rights Violation}:
For the task prompts (Table \ref{tab:HR_task_prompt_gemini} and \ref{tab:HR_task_prompt_Llama}) the LLMs have worse Macro-F1 scores than the baseline Legal-BERT. The better performance of Legal-BERT (despite having lower token limit of 512 tokens) could be due to the fact that it is already pre-trained on ECHR corpus. After using the LLMs to `run' the ReQuest algorithm (obtained through similar prompting methods as stated in the statute prediction task), using robustness check prompts (Tables \ref{tab:robustness_HR_gemini} and \ref{tab:robustness_HR_Llama} in the Appendix) we obtained very high PeRR and PreRR scores for {\it{intra-LLM}} setup for both \texttt{gemini} and \texttt{llama3} as compared to the statute prediction task which has a considerably larger test data. Similar trends were observed for the {\it{inter-LLM}} setups as well, although the scores were a bit lower than that of {\it{intra-LLM}} setup.

\begin{table*}[h]

\centering
\makebox[\textwidth][c]{ 
\rowcolors{1}{tablebg}{white}
\tiny
\begin{tabular}{|c|c|c|c|c|c|c|c|c|c|}
\hline
\rowcolor{lightred}
\multicolumn{10}{|c|}{Apple (AAPL)} \\ 
\hline
\multicolumn{5}{|c|}{Intra-LLM} & \multicolumn{5}{|c|}{Inter-LLM} \\ 
\hline
\multicolumn{3}{|c|}{Macro F1} &  Percentage & Ratio & Macro F1 &  Percentage & Ratio &  Percentage & Ratio\\ 
\hline
{\bf baseline (BERTweet)} & {\bf Gemini} & {\bf ReQuest algo} &  {\bf PerRR\_GP\_G} & {\bf PreRR\_GP\_G} & {\bf Gemini $\rightarrow$ Llama} & {\bf PerRR\_GP\_L} & {\bf PreRR\_GP\_L} & {\bf PerRR\_GA\_L} & {\bf PreRR\_GA\_L}\\
\hline
                   0.53            & 0.44        &  0.48       & 91 {\color{green}$\uparrow$} & 0.6458 & 0.48 & 91.67 &  0.4791 & 100{\color{green}$\uparrow$} & 0.5520\\
\hline                   
 {\bf baseline (BERTweet)} & {\bf Llama} & {\bf ReQuest algo} &  {\bf PerRR\_LP\_L} & {\bf PreRR\_LP\_L} & {\bf Llama $\rightarrow$ Gemini} & {\bf PerRR\_LP\_G} & {\bf PreRR\_LP\_G} & {\bf PerRR\_LA\_G} & {\bf PreRR\_LA\_G}\\
\hline
 0.53             & 0.41       & 0.38       & 92.68 {\color{red}$\downarrow$} & 0.9375 & 0.52 & 78.84 &  0.4979 & 78.84 {\color{green}$\uparrow$} & 0.4916\\
\hline

\rowcolor{lightred}
\multicolumn{10}{|c|}{Microsoft (MSFT)} \\ 
\hline
\multicolumn{5}{|c|}{Intra-LLM} & \multicolumn{5}{|c|}{Inter-LLM} \\ 
\hline
\multicolumn{3}{|c|}{Macro F1} &  Percentage & Ratio & Macro F1 &  Percentage & Ratio &  Percentage & Ratio\\ 
\hline
{\bf Baseline (BERTweet)} & {\bf Gemini} & {\bf ReQuest algo} &  {\bf PerRR\_GP\_G} & {\bf PreRR\_GP\_G} & {\bf Gemini $\rightarrow$ Llama} & {\bf PerRR\_GP\_L} & {\bf PreRR\_GP\_L} & {\bf PerRR\_GA\_L} & {\bf PreRR\_GA\_L}\\
\hline
0.51 & 0.42 & 0.51 & 82 {\color{red}$\downarrow$} & 0.5064 & 0.44 & 95.45 & 0.375 & 86.27{\color{green}$\uparrow$} & 0.6314\\
\hline                   
{\bf Baseline (BERTweet)} & {\bf Llama} & {\bf ReQuest algo} &  {\bf PerRR\_LP\_L} & {\bf PreRR\_LP\_L} & {\bf Llama $\rightarrow$ Gemini} & {\bf PerRR\_LP\_G} & {\bf PreRR\_LP\_G} & {\bf PerRR\_LA\_G} & {\bf PreRR\_LA\_G}\\
\hline
0.51 & 0.41 &  0.37 & 90.2 {\color{red}$\downarrow$} & 0.5021 & 0.50 & 91.11 & 0.4482 & 74{\color{red}$\downarrow$} & 0.3987\\
\hline

\end{tabular}
} 

\caption{Reproducibility: Finance. The results show good to medium reproducibility performance on PerRR and medium performance on PreRR, the latter showcasing data-level variability of LLMs.}
\label{tab:finance}
\end{table*}

\section{Finance Tasks}\label{sec:finance}



Investors often consider social media when making decisions, as sentiment analysis captures information not yet reflected in prices. In this study, our objective is to predict upward or downward movement of stocks based on tweets from preceding two days. Yahoo! finance \footnote{\url{https://finance.yahoo.com/sectors/}}  classifies stocks into eleven distinct industry sectors. Given that stocks with high trading volumes are frequently mentioned on Twitter, we have chosen to focus on the price movements of two stocks, from a pool of 88, over the period from January 1, 2014, to January 1, 2016 \cite{xu-cohen-2018-stock}
\cite{sawhney-etal-2020-deep}.

We focus on \noindent {\bf binary classification } to determine price changes as influenced by social media sentiment. A positive sentiment around the stock indicates a potential upward movement~(labeled as 1), whereas a negative sentiment anticipates a downward movement, labeled as 0.

The tweets, encompassing over 20 mentions of a particular stock, are fed into the LLM with a task prompt ($\mathcal{T}$). The LLM is instructed to predict the movement of the stock based on these given tweets. The LLM analyzes the sentiment and content of the tweets to determine whether the stock price may move up or down. Subsequently, a ReQuest prompt ($\mathcal{R}$) is employed as a follow-up to extract the ReQuest algorithm ($\mathcal{A}$) that ensures reproducibility. This algorithm ($\mathcal{A}$) is used for both {\it{intra-LLM}} and {\it{inter-LLM}} setups, as with previous datasets. 

\subsection{Results}
For our baseline, we have chosen {\bf BERTweet}~\cite{BERTweet} for sentiment analysis. In this setup, tweets are directly fed into BERTweet to generate sentiment scores. These sentiment scores are then used to predict whether the stock price will move up or down. This approach provides a foundational benchmark against which we can compare the performance of LLMs and the {\it{ReQuest algorithm}}. Details of the results are available in Appendix \ref{app:finance}.

Our experiments on the reproducibility of language models (LLMs) using the {\it{ReQuest algorithm}} revealed key trends across {\it{intra}}- and {\it{inter-LLM}} setups for Apple (AAPL) and Google (GOOG) datasets. For instance, in the {\it{intra-LLM}} setup for Apple (AAPL), the baseline Macro F1 score with BERTweet was 0.53. \texttt{Gemini}'s Macro F1 score improved from 0.44 to 0.48 with ($\mathcal{A}$) , achieving a high reproducibility percentage (PerRR\_GP\_G) of 91\%. Conversely, \texttt{llama3}'s Macro F1 score decreased from 0.41 to 0.38 with ($\mathcal{A}$), despite a high reproducibility (PerRR\_LP\_L) of 92.68\%. In the {\it{inter-LLM}} setup for Apple (AAPL), when ($\mathcal{A}$) generated by \texttt{gemini} was applied to \texttt{llama3}, the PerRR\_GP\_L was 91.67\%, indicating a high degree of reproducibility, though the Macro F1 score was moderately lower at 0.48. Conversely, applying ($\mathcal{A}$) generated by \texttt{llama3} to \texttt{gemini} resulted in a PerRR\_LP\_G of 78.84\%, indicating lower reproducibility and a Macro F1 score of 0.52.
High {\it{intra-LLM}} reproducibility for both \texttt{gemini} and \texttt{llama3} indicates consistent performance within the same model.
 {\it{Inter-LLM}} reproducibility is variable, with \texttt{gemini-1.0-pro} showing better transferability to \texttt{llama3} than vice versa. Although \texttt{llama3}'s accuracy was lower, it exhibited higher PreRR values, indicating better prediction-level reproducibility. This suggests that \texttt{llama3}'s predictions were more consistent at the data level, despite its overall lower performance compared to \texttt{gemini}.

\begin{table*}[h]
\centering
\makebox[\textwidth][c]{ 
\rowcolors{1}{tablebg}{white}
\scriptsize
\begin{tabular}{|c|c|c|c|c|c|c|c|c|c|}
\hline
\rowcolor{lightred}
\multicolumn{10}{|c|}{Reddit Suicide Watch Dataset~\citep{ji2021suicidal} } \\ 
\hline
\multicolumn{5}{|c|}{Intra-LLM} & \multicolumn{5}{|c|}{Inter-LLM} \\ 
\hline
\multicolumn{3}{|c|}{Macro F1} &  Percentage & Ratio & Macro F1 &  Percentage & Ratio &  Percentage & Ratio\\ 
\hline
{\bf baseline (MentalBERT)} & {\bf Gemini} & {\bf ReQuest algo} &  {\bf PerRR\_GP\_G} & {\bf PreRR\_GP\_G} & {\bf Gemini $\rightarrow$ Llama} & {\bf PerRR\_GP\_L} & {\bf PreRR\_GP\_L} & {\bf PerRR\_GA\_L} & {\bf PreRR\_GA\_L}\\
\hline
                   0.7111           & 0.6363         &  0.5761        & 90.54 {\color{red}$\downarrow$} & 0.617 & 0.6363 & 95.47 {\color{green}$\uparrow$} &  0.728 & 86.47 {\color{green}$\uparrow$} & 0.731\\
\hline                   
 {\bf baseline (MentalBERT)} & {\bf Llama} & {\bf ReQuest algo} &  {\bf PerRR\_LP\_L} & {\bf PreRR\_LP\_L} & {\bf Llama $\rightarrow$ Gemini} & {\bf PerRR\_LP\_G} & {\bf PreRR\_LP\_G} & {\bf PerRR\_LA\_G} & {\bf PreRR\_LA\_G}\\
\hline
 0.7111             & 0.5789       & 0.6136         & 94.34  {\color{green}$\uparrow$} & 0.701 & 0.5789  & 99.31 {\color{red}$\downarrow$} &  0.583 & 94.99 {\color{green}$\uparrow$} & 0.713\\
 \hline
\hline
\rowcolor{lightred}
\multicolumn{10}{|c|}{Depression Severity Dataset~\citep{naseem2022early} } \\ 
\hline
\multicolumn{5}{|c|}{Intra-LLM} & \multicolumn{5}{|c|}{Inter-LLM} \\ 
\hline
\multicolumn{3}{|c|}{Macro F1} &  Percentage & Ratio & Macro F1 &  Percentage & Ratio &  Percentage & Ratio\\ 
\hline
{\bf baseline (L-BERT)} & {\bf Gemini} & {\bf ReQuest algo} &  {\bf PerRR\_GP\_G} & {\bf PreRR\_GP\_G} & {\bf Gemini $\rightarrow$ Llama} & {\bf PerRR\_GP\_L} & {\bf PreRR\_GP\_L} & {\bf PerRR\_GA\_L} & {\bf PreRR\_GA\_L}\\
\hline
 0.5556           & 0.29147       &  0.28417       & 97.545 {\color{red}$\downarrow$} & 0.672 & 0.29147                           & 86.70 {\color{red}$\downarrow$} &  0.657 & 88.96 {\color{red}$\downarrow$} & 0.605\\
\hline                   
 {\bf baseline (L-BERT)} & {\bf Llama} & {\bf ReQuest algo} &  {\bf PerRR\_LP\_L} & {\bf PreRR\_LP\_L} & {\bf Llama $\rightarrow$ Gemini} & {\bf PerRR\_LP\_G} & {\bf PreRR\_LP\_G} & {\bf PerRR\_LA\_G} & {\bf PreRR\_LA\_G}\\
\hline
0.5556           & 0.29728       & 0.30117        & 98.7  {\color{green}$\uparrow$} & 0.986 & 0.29728 & 87.88 {\color{red}$\downarrow$} & 0.5786 & 86.74 {\color{red}$\downarrow$} & 0.575\\
\hline
\end{tabular}
} 

\caption{Reproducibility: Health. The results show high to medium reproducibility performance on PerRR and medium performance on PreRR, the latter showcasing data-level variability of LLMs. Specifically, on PerRR, the values are sometimes very close to 100, exhibiting high degree of reproducibility of the ReQuest algorithm.
}
\label{tab:health}
\end{table*}

\section{Health tasks}\label{sec:health}

Early detection of mental disorders and suicidal tendencies from digital footprint has become increasingly crucial, providing an opportunity for timely intervention. We select a dataset for the task of detection from social-media posts on Reddit, $T_1$, \cite{ji2021suicidal} with 5 classes $C_1$ covering five different types of psychological ailment.
Classifying suicidal tendencies and related disorders is challenging due to overlapping semantics. We therefore aim to uncover the inner black-box mechanism of an LLM as it tackles this complex task. 
Table \ref{tab:disease_task_prompt} shows the zero-shot classification prompt.
Upon getting the initial classifications, we provide them as examples in the ReQuest prompt ($\mathcal{R}$) to the LLM in order to extract the ReQuest algorithm ($\mathcal{A}$) (refer Table \ref{tab:disease_algo}) for the task. The algorithm, in turn, is used in the Robustness Check Prompt (refer Table \ref{tab:robusteness_disease}) to get another set of classifications from the LLM.\\
We consider another dataset on Reddit posts~\cite{naseem2022early} involving detection of the severity of depression, where each text is classified into one of 4 levels of severity~ $C_2$. 
We utilize the task prompt ($\mathcal{T}$)~(Table \ref{tab:medicine_task_prompt}) to prompt the candidate LLMs -- \texttt{gemini-1.0-pro} and \texttt{llama3} and obtain a set of zero-shot classifications. 
The ReQuest algorithm $\mathcal{A}$ (Table \ref{tab:medicine_algo}) is extracted with the help of $\mathcal{R}$. Subsequently, the Robustness Check prompt, embedded with $\mathcal{A}$, is used to get LLM responses to test the viability of the ReQuest algorithm. Refer to Table \ref{tab:robustness_medicine} for details of the Robustness Check prompt. No significant performance differences were observed across multiple runs.

\subsection{Prompting}
The LLM is assigned the role of a medical professional specializing in mental health and depression detection, an expert at detecting suicidal tendencies (for the task $T_1$)
and classifying the depression severity (for task $T_2$) from a given text. We follow this up with its task—through extensive iterative prompt development—to make the classifications exact matches. We repeatedly instruct the LLM to neglect any extra information other than the class name, to avert possibility of hallucinations.
Despite our efforts, there are cases of anomalies which may be due to hallucinations by LLM. Table \ref{tab:disease_task_prompt} provides an example of the task prompt ($\mathcal{T}$) used for \texttt{gemini}.

\subsection{Results}

We opt for macro-avg F1 scores as the evaluation metric due to the highly imbalanced nature of the datasets. To establish a baseline, we fine-tune mental-BERT~\cite{ji2021mentalbertpubliclyavailablepretrained} for tasks $T_1$ and $T_2$ respectively, and compare the zero-shot classification performance of the candidate LLMs against the reported values. 
In Table \ref{tab:health}, we observe that both \texttt{gemini} and \texttt{llama3} perform relatively better in $T_1$. 
\newline
PreRR values are notably lower than PerRR values across the board. This indicates that even when the overall performance (as measured by Macro F1 and PerRR) is similar, the actual predictions performed by the LLMs using the generated algorithms ($\mathcal{A}$) often differ considerably from the initial predictions. This would suggest \textbf{lack of deterministic reasoning at the individual data point level}.
However, for \texttt{llama3}, the responses using Robustness Check prompt involving the ReQuest algorithm $\mathcal{A}$ appears to be better than Task prompt responses.
Further, $\mathcal{A}$ from \texttt{llama3} performs better than that from 
\texttt{gemini} in both Intra and Inter LLM setuprs for $T_1$. For $T_2$, it is better for Intra-LLM setup but not in the Inter-LLM setup. Refer to Section \ref{app:health} for a detailed analysis of the tasks.

\section{RQ3: Does the ReQuest algorithm align with the ``intrinsic reasoning'' of the LLM?}\label{sec:intrinsic}
Our work has so far attempted to extract a \textit{faithful} representation of the candidate model's reasoning in the form of a human-verifiable algorithm, through manual refinement of the ReQuest prompts. The faithfulness of the possible representations, in our work, is decided through the reproducibility of results. Now we ponder upon a deeper question: \textit{Can we determine faithfulness in a different manner, considering the process of answer generation?} \citet{CoTRWOP} reveals that pretrained language models \textit{inherently posses reasoning capabilities} which requires no prompting to elicit. A language model follows next token prediction and based on the most probable token at each step a sequence is formed. The approach considers top-k tokens \{$t_1,...t_k$\} and for $t_i$ generates a sequence on greedy basis from the second token~(step), exploring k decoding paths in total. The work explores alternative decoding paths in the model based on the top-k tokens at the first level and shows that a high-confidence decoding path often generates the reasoning leading to a more reliable answer compared to a traditional greedy decoding method, which may directly yield an unreliable answer through a low-confidence path.~A major limitation of the approach is closed-source models where the generation process cannot be emulated. We adopt the approach in our work to elicit reasoning or explanation from the model for classification and attempt to draw equivalence to the LLM-yielded human-readable algorithm in terms of categorization of an instance.

We employ LLaMA 3.2-1B (Instruct) for inference, executing the original prompts initially. 
To account for the considerable parameter size difference, we iteratively made minor changes to ensure the task is correctly communicated to a comparatively smaller model, without altering the structure of the prompt. We limit the exploration to k=10. 
\begin{table}[]
    \centering
    \small
    \begin{tabular}{|p{7cm}|}
    \hline
    \textbf{ReQuest Algo}\\
    ...\\
    Step 2: \colorbox{green!30}{Feature Extraction}\\
    ...Sequences of adjacent tokens (e.g., bigrams or trigrams) can be extracted to \colorbox{green!30}{capture phrases or expressions that convey specific} \colorbox{green!30}{sentiments or topics}...\\
    \hline
    \textbf{Intrinsic Reasoning}\\
    ...The \colorbox{green!30}{key phrases in the text} are "wanting to skip or postpone my exam", "my exam is on saturday", "i feel a massive urge to reschedule", "i experience this often with major exams", "i just dont feel ready", "i dont know im so tired and mentally fried", "a part of me thinks i should fight this urge and just tank it"...The text has a \colorbox{green!30}{negative emotional tone}, with phrases such as "i feel a massive urge to reschedule", "i just dont feel ready"...the correct class of the text is "self-Anxiety"...\\
    \hline
    \end{tabular}
    \caption{Example of step-wise \colorbox{green!30}{alignment} between the LLM-returned ReQuest algorithm and the reasoning intrinsic to the generated classification response. We observe execution correspondence indicating a possible avenue of verification.}
    \label{Intrinsic-Reason}
\end{table}

Table~\ref{Intrinsic-Reason} provides a juxtaposition of the reasoning elicited through the approach versus parts of the ReQuest algorithm that closely correspond to that intrinsic reasoning. A major difference between the model-generated ReQuest algorithm and the reasoning embedded in the sequence generated via no explicit instruction in the prompt is the presentation of steps. The ReQuest algorithm entails additional set-up steps e.g. a preprocessing step is recommended, which resembles a pipeline of execution. In the absence of an established method, we manually verify and interpret the sequence-based reasoning and compare it with the ReQuest algorithm for correspondence (Table \ref{Intrinsic-Reason}). Considering the promising results, we aim to further study in the future the potential of the approach in our work. 


\section{Conclusion}

 Generally, high reproducibility is exhibited in case of legal tasks, particularly in the intra-LLM setup. This may suggest that LLMs are capable of generating algorithms that can closely mimic their initial responses when applied to the same model.

\section*{Limitations}
We have mainly restricted our study to {\it zero-shot} prompting throughout this work, due to strict token
limits. 
For \texttt{gemini}, certain data points were blocked due to safety settings and recitation errors. In the medical dataset, some words triggered \texttt{gemini}'s safety mechanisms, and similar issues arose with certain tweets in the finance dataset. 

Due to limited resources, we utilized a notably smaller version of \texttt{llama3} to study intrinsic reasoning in the LLM, the performance of which may not be comparable to the candidate LLMs prompted throughout the work. We however plan to repeat the process in a comparable-sized LLM upon availability of adequate computational machinary.

{
We acknowledge that factors like the prompt template and context may influence reproducibility. But our objective was to enhance robustness without controlling for variables typically beyond our control in real-world applications. The high reproducibility values observed in legal and financial contexts show that the ReQuesting framework effectively improves reproducibility in LLMs despite the lack of control over other influencing factors.
}

\bibliography{main}

\begin{thebibliography}{57}
\providecommand{\natexlab}[1]{#1}

\bibitem[{Jen()}]{Jenner}

\newblock \href {https://www.bbc.co.uk/bitesize/guides/zxbqjsg/revision/6} {\url{https://www.bbc.co.uk/bitesize/guides/zxbqjsg/revision/6}}.

\bibitem[{Band et~al.(2024)Band, Li, Ma, and Hashimoto}]{band2024linguistic}
Neil Band, Xuechen Li, Tengyu Ma, and Tatsunori Hashimoto. 2024.
\newblock \href {https://arxiv.org/abs/2404.00474} {Linguistic calibration of long-form generations}.
\newblock \emph{Preprint}, arXiv:2404.00474.

\bibitem[{Blackwell(1991)}]{Galileo}
Richard~J Blackwell. 1991.
\newblock Galileo, bellarmine, and the bible.
\newblock \emph{University of Notre Dame Press}.

\bibitem[{Cascella et~al.(2023)Cascella, Montomoli, Bellini, and Bignami}]{Cascella2023EvaluatingTF}
M.~Cascella, J.~Montomoli, V.~Bellini, and E.~Bignami. 2023.
\newblock \href {https://doi.org/10.1007/s10916-023-01925-4} {Evaluating the feasibility of chatgpt in healthcare: An analysis of multiple clinical and research scenarios}.
\newblock \emph{Journal of Medical Systems}, 47(1):33.

\bibitem[{Chalkidis et~al.(2019)Chalkidis, Androutsopoulos, and Aletras}]{chalkidis-etal-2019-neural}
Ilias Chalkidis, Ion Androutsopoulos, and Nikolaos Aletras. 2019.
\newblock \href {https://doi.org/10.18653/v1/P19-1424} {Neural legal judgment prediction in {E}nglish}.
\newblock In \emph{Proceedings of the 57th Annual Meeting of the Association for Computational Linguistics}, pages 4317--4323, Florence, Italy. Association for Computational Linguistics.

\bibitem[{Chalkidis et~al.(2020)Chalkidis, Fergadiotis, Malakasiotis, Aletras, and Androutsopoulos}]{chalkidis-etal-2020-legal}
Ilias Chalkidis, Manos Fergadiotis, Prodromos Malakasiotis, Nikolaos Aletras, and Ion Androutsopoulos. 2020.
\newblock \href {https://doi.org/10.18653/v1/2020.findings-emnlp.261} {{LEGAL}-{BERT}: The muppets straight out of law school}.
\newblock In \emph{Findings of the Association for Computational Linguistics: EMNLP 2020}, pages 2898--2904, Online. Association for Computational Linguistics.

\bibitem[{Chen et~al.(2023)Chen, Wang, Long, Zhang, Lu, Li, Wang, Xu, Bai, Huang, and Wei}]{chen2023discfinllm}
Wei Chen, Qiushi Wang, Zefei Long, Xianyin Zhang, Zhongtian Lu, Bingxuan Li, Siyuan Wang, Jiarong Xu, Xiang Bai, Xuanjing Huang, and Zhongyu Wei. 2023.
\newblock \href {https://arxiv.org/abs/2310.15205} {Disc-finllm: A chinese financial large language model based on multiple experts fine-tuning}.
\newblock \emph{Preprint}, arXiv:2310.15205.

\bibitem[{Chen et~al.(2024)Chen, Xiang, Lu, Liu, He, and Shi}]{chen2024evaluating}
Xiaolan Chen, Jiayang Xiang, Shanfu Lu, Yexin Liu, Mingguang He, and Danli Shi. 2024.
\newblock \href {https://arxiv.org/abs/2405.07468} {Evaluating large language models in medical applications: a survey}.
\newblock \emph{Preprint}, arXiv:2405.07468.

\bibitem[{De~Angelis et~al.(2023)De~Angelis, Baglivo, Arzilli, Privitera, Ferragina, Tozzi, and Rizzo}]{deangelis2023chatgpt}
L.~De~Angelis, F.~Baglivo, G.~Arzilli, G.~P. Privitera, P.~Ferragina, A.~E. Tozzi, and C.~Rizzo. 2023.
\newblock \href {https://doi.org/10.3389/fpubh.2023.1166120} {Chatgpt and the rise of large language models: the new ai-driven infodemic threat in public health}.
\newblock \emph{Front. Public Health}, 11.

\bibitem[{Deroy et~al.(2023)Deroy, Ghosh, and Ghosh}]{ghosh2023}
Aniket Deroy, Kripabandhu Ghosh, and Saptarshi Ghosh. 2023.
\newblock \href {https://ceur-ws.org/Vol-3423/} {How ready are pre-trained abstractive models and llms for legal case judgement summarization?}
\newblock \emph{Proceedings of the Third International Workshop on Artificial Intelligence and Intelligent Assistance for Legal Professionals in the Digital Workplace}.

\bibitem[{Franc et~al.(2024)Franc, Cheng, Hart, Yadav, and Yadav}]{Franc2024}
J.M. Franc, L.~Cheng, A.~Hart, K.~Yadav, and K.~Yadav. 2024.
\newblock \href {https://doi.org/10.1007/s43678-023-00616-w} {Repeatability, reproducibility, and diagnostic accuracy of a commercial large language model (chatgpt) to perform emergency department triage using the canadian triage and acuity scale}.
\newblock \emph{Canadian Journal of Emergency Medicine}.

\bibitem[{Herrmann and Levinstein(2024)}]{herrmann2024standards}
Daniel~A. Herrmann and Benjamin~A. Levinstein. 2024.
\newblock \href {https://arxiv.org/abs/2405.21030} {Standards for belief representations in llms}.
\newblock \emph{Preprint}, arXiv:2405.21030.

\bibitem[{Hobsbawm(1962)}]{IndustrRev}
Eric Hobsbawm. 1962.
\newblock The age of revolution: Europe 1789–1848.
\newblock \emph{New York, New American Library}.

\bibitem[{Huang et~al.(2023{\natexlab{a}})Huang, Mamidanna, Jangam, Zhou, and Gilpin}]{huang2023largelanguagemodelsexplain}
Shiyuan Huang, Siddarth Mamidanna, Shreedhar Jangam, Yilun Zhou, and Leilani~H. Gilpin. 2023{\natexlab{a}}.
\newblock \href {https://arxiv.org/abs/2310.11207} {Can large language models explain themselves? a study of llm-generated self-explanations}.
\newblock \emph{Preprint}, arXiv:2310.11207.

\bibitem[{Huang et~al.(2023{\natexlab{b}})Huang, Zhang, Y, and Sun}]{huang2023trustgpt}
Yue Huang, Qihui Zhang, Philip~S. Y, and Lichao Sun. 2023{\natexlab{b}}.
\newblock \href {https://arxiv.org/abs/2306.11507} {Trustgpt: A benchmark for trustworthy and responsible large language models}.
\newblock \emph{Preprint}, arXiv:2306.11507.

\bibitem[{Jaccard(1912)}]{Jaccard}
Paul Jaccard. 1912.
\newblock \href {https://doi.org/10.1111/j.1469-8137.1912.tb05611.x} {The distribution of the flora in the alpine zone.1}.
\newblock \emph{New Phytologist}, 11(2):37--50.

\bibitem[{Ji et~al.(2021{\natexlab{a}})Ji, Li, Huang, and Cambria}]{ji2021suicidal}
Shaoxiong Ji, Xue Li, Zi~Huang, and Erik Cambria. 2021{\natexlab{a}}.
\newblock Suicidal ideation and mental disorder detection with attentive relation networks.
\newblock \emph{Neural Computing and Applications}.

\bibitem[{Ji et~al.(2021{\natexlab{b}})Ji, Zhang, Ansari, Fu, Tiwari, and Cambria}]{ji2021mentalbertpubliclyavailablepretrained}
Shaoxiong Ji, Tianlin Zhang, Luna Ansari, Jie Fu, Prayag Tiwari, and Erik Cambria. 2021{\natexlab{b}}.
\newblock \href {https://arxiv.org/abs/2110.15621} {Mentalbert: Publicly available pretrained language models for mental healthcare}.
\newblock \emph{Preprint}, arXiv:2110.15621.

\bibitem[{Karabacak and Margetis(2023)}]{Karabacak2023embracing}
Mert Karabacak and Konstantinos Margetis. 2023.
\newblock \href {https://doi.org/10.7759/cureus.39305} {Embracing large language models for medical applications: Opportunities and challenges}.
\newblock \emph{Cureus}, 15(5):e39305.

\bibitem[{Katz et~al.(2023)Katz, Bommarito, Gao, and Arredondo}]{katz2023}
Daniel~Martin Katz, Michael~James Bommarito, Shang Gao, and Pablo Arredondo. 2023.
\newblock \href {https://doi.org/10.2139/ssrn.4389233} {Gpt-4 passes the bar exam}.
\newblock \emph{Philosophical Transactions of the Royal Society A}.

\bibitem[{Khan and Raza(2021)}]{Khan2021}
K.~J. Khan and V.~F. Raza. 2021.
\newblock \href {https://doi.org/10.1136/bcr-2020-235542} {Specialist shortage in developing countries: comprehending delays in care}.
\newblock \emph{BMJ Case Reports}, 14(1):e235542.

\bibitem[{Kim et~al.(2024{\natexlab{a}})Kim, Muhn, and Nikolaev}]{kim2024financial}
Alex~G. Kim, Maximilian Muhn, and Valeri~V. Nikolaev. 2024{\natexlab{a}}.
\newblock \href {https://ssrn.com/abstract=4835311} {Financial statement analysis with large language models}.

\bibitem[{Kim et~al.(2024{\natexlab{b}})Kim, Isozaki, Sirkin, and Robson}]{kim2024generativeartificialintelligencereproducibility}
Edward Kim, Isamu Isozaki, Naomi Sirkin, and Michael Robson. 2024{\natexlab{b}}.
\newblock \href {https://arxiv.org/abs/2307.01898} {Generative artificial intelligence reproducibility and consensus}.
\newblock \emph{Preprint}, arXiv:2307.01898.

\bibitem[{Lai et~al.(2023)Lai, Gan, Wu, Qi, and Yu}]{lai2023large}
Jinqi Lai, Wensheng Gan, Jiayang Wu, Zhenlian Qi, and Philip~S. Yu. 2023.
\newblock \href {https://arxiv.org/abs/2312.03718} {Large language models in law: A survey}.
\newblock \emph{Preprint}, arXiv:2312.03718.

\bibitem[{Lakkaraju et~al.(2023)Lakkaraju, Vuruma, Pallagani, Muppasani, and Srivastava}]{lakkaraju2023llms}
Kausik Lakkaraju, Sai Krishna~Revanth Vuruma, Vishal Pallagani, Bharath Muppasani, and Biplav Srivastava. 2023.
\newblock \href {https://arxiv.org/abs/2307.07422} {Can llms be good financial advisors?: An initial study in personal decision making for optimized outcomes}.
\newblock \emph{Preprint}, arXiv:2307.07422.

\bibitem[{Liu et~al.(2023)Liu, Xia, Wang, and Zhang}]{codeNeurips2023}
Jiawei Liu, Chunqiu~Steven Xia, Yuyao Wang, and Lingming Zhang. 2023.
\newblock \href {https://proceedings.neurips.cc/paper_files/paper/2023/file/43e9d647ccd3e4b7b5baab53f0368686-Paper-Conference.pdf} {Is your code generated by chatgpt really correct? rigorous evaluation of large language models for code generation}.
\newblock In \emph{Advances in Neural Information Processing Systems}, volume~36, pages 21558--21572. Curran Associates, Inc.

\bibitem[{Most et~al.(2024)Most, Hu, Yang, Liu, Chen, Li, Xu, Liu, and Sikora}]{Most2024.03.21.24304667}
Amoreena Most, Mengxuan Hu, Huibo Yang, Tianming Liu, Xianyan Chen, Sheng Li, Steven Xu, Zhengliang Liu, and Andrea Sikora. 2024.
\newblock \href {https://doi.org/10.1101/2024.03.21.24304667} {Evaluating accuracy and reproducibility of large language model performance in pharmacy education}.
\newblock \emph{medRxiv}.

\bibitem[{Naseem et~al.(2022)Naseem, Dunn, Kim, and Khushi}]{naseem2022early}
Usman Naseem, Adam~G Dunn, Jinman Kim, and Matloob Khushi. 2022.
\newblock Early identification of depression severity levels on reddit using ordinal classification.
\newblock In \emph{Proceedings of the ACM Web Conference 2022}, pages 2563--2572.

\bibitem[{Nguyen et~al.(2020)Nguyen, Vu, and Tuan~Nguyen}]{BERTweet}
Dat~Quoc Nguyen, Thanh Vu, and Anh Tuan~Nguyen. 2020.
\newblock \href {https://doi.org/10.18653/v1/2020.emnlp-demos.2} {{BERT}weet: A pre-trained language model for {E}nglish tweets}.
\newblock In \emph{Proceedings of the 2020 Conference on Empirical Methods in Natural Language Processing: System Demonstrations}, pages 9--14, Online. Association for Computational Linguistics.

\bibitem[{Paul et~al.(2024)Paul, Bhatt, Goyal, and Ghosh}]{10.1145/3626772.3657879}
Shounak Paul, Rajas Bhatt, Pawan Goyal, and Saptarshi Ghosh. 2024.
\newblock \href {https://doi.org/10.1145/3626772.3657879} {Legal statute identification: A case study using state-of-the-art datasets and methods}.
\newblock In \emph{Proceedings of the 47th International ACM SIGIR Conference on Research and Development in Information Retrieval}, SIGIR '24, page 2231–2240, New York, NY, USA. Association for Computing Machinery.

\bibitem[{Rae et~al.(2023)Rae, Borgeaud, Cai, Millican, Hoffmann, Song, Aslanides, Henderson, Ring, Young, Rutherford, Hennigan, Menick, Cassirer, Powell, van~den Driessche, Hendricks, Rauh, Huang, Spreeuwenberg, Agarwal, Lacroix, Green, Cowie, Harley, Ballard, Sifre, Kavukcuoglu, Rockt{"a}schel, Pondard, Adler, Chowdhary, Ramos, Saunders, Hayman, Jumper, Kohli, Nematzadeh, Reynolds, Saxton, Green, Noland, Askell, Yogatama, Cohen, McKinney, Smith, Hassabis, Kavukcuoglu, Kohli, and Irving}]{Rae2023}
Jack~W. Rae, Sebastian Borgeaud, Tim Cai, Katie Millican, Jonas Hoffmann, Francis Song, John Aslanides, Sarah Henderson, Roman Ring, Susannah Young, Eliza Rutherford, Tom Hennigan, Jacob Menick, Albin Cassirer, Richard Powell, George van~den Driessche, Lisa~Anne Hendricks, Maribeth Rauh, Po-Sen Huang, Amelia Spreeuwenberg, Michal Agarwal, Arthur Lacroix, Susie Green, Irina Cowie, Cyril Harley, Andrew Ballard, Leo Sifre, Koray Kavukcuoglu, Tim Rockt{"a}schel, Nicolas Pondard, Tom Adler, Kirsty~R. Chowdhary, Serkan Ramos, Dominic~M. Saunders, Joseph Hayman, John Jumper, Pushmeet Kohli, Alison Nematzadeh, Fabian Reynolds, David Saxton, Emily Green, Luc Noland, Amanda Askell, Dani Yogatama, Katja Cohen, Susannah McKinney, Owen Smith, Demis Hassabis, Koray Kavukcuoglu, Pushmeet Kohli, and Geoffrey Irving. 2023.
\newblock \href {https://doi.org/10.1038/s41586-023-06291-2} {Large language models encode clinical knowledge}.
\newblock \emph{Nature}, 620(7897):172--180.

\bibitem[{Ribeiro et~al.(2016)Ribeiro, Singh, and Guestrin}]{lime2016}
Marco~Tulio Ribeiro, Sameer Singh, and Carlos Guestrin. 2016.
\newblock \href {https://doi.org/10.1145/2939672.2939778} {"why should i trust you?": Explaining the predictions of any classifier}.
\newblock In \emph{Proceedings of the 22nd ACM SIGKDD International Conference on Knowledge Discovery and Data Mining}, KDD '16, page 1135–1144, New York, NY, USA. Association for Computing Machinery.

\bibitem[{Rudin et~al.(2020)Rudin, Wang, and Coker}]{compas}
Cynthia Rudin, Caroline Wang, and Beau Coker. 2020.
\newblock The {Age} of {Secrecy} and {Unfairness} in {Recidivism} {Prediction}.
\newblock \emph{Harvard Data Science Review}, 2(1).
\newblock Https://hdsr.mitpress.mit.edu/pub/7z10o269.

\bibitem[{Sarker(2024)}]{sarker2024llm}
I.H. Sarker. 2024.
\newblock \href {https://doi.org/10.1007/s44163-024-00129-0} {Llm potentiality and awareness: a position paper from the perspective of trustworthy and responsible ai modeling}.
\newblock \emph{Discov Artif Intell}, 4:40.

\bibitem[{Sarker et~al.(2024)Sarker, Downing, Desai, and Bultan}]{sarker2024syntacticrobustnessllmbasedcode}
Laboni Sarker, Mara Downing, Achintya Desai, and Tevfik Bultan. 2024.
\newblock \href {https://arxiv.org/abs/2404.01535} {Syntactic robustness for llm-based code generation}.
\newblock \emph{Preprint}, arXiv:2404.01535.

\bibitem[{Sawhney et~al.(2020)Sawhney, Agarwal, Wadhwa, and Shah}]{sawhney-etal-2020-deep}
Ramit Sawhney, Shivam Agarwal, Arnav Wadhwa, and Rajiv~Ratn Shah. 2020.
\newblock \href {https://doi.org/10.18653/v1/2020.emnlp-main.676} {Deep attentive learning for stock movement prediction from social media text and company correlations}.
\newblock In \emph{Proceedings of the 2020 Conference on Empirical Methods in Natural Language Processing (EMNLP)}, pages 8415--8426, Online. Association for Computational Linguistics.

\bibitem[{Sclar et~al.(2024)Sclar, Choi, Tsvetkov, and Suhr}]{sclar2024quantifyinglanguagemodelssensitivity}
Melanie Sclar, Yejin Choi, Yulia Tsvetkov, and Alane Suhr. 2024.
\newblock \href {https://arxiv.org/abs/2310.11324} {Quantifying language models' sensitivity to spurious features in prompt design or: How i learned to start worrying about prompt formatting}.
\newblock \emph{Preprint}, arXiv:2310.11324.

\bibitem[{Stade et~al.(2024)Stade, Stirman, Ungar, Boland, Schwartz, Yaden, Sedoc, DeRubeis, Willer, and Eichstaedt}]{healthtrustLLM}
Elizabeth~C. Stade, Shannon~Wiltsey Stirman, Lyle~H. Ungar, Cody~L. Boland, H.~Andrew Schwartz, David~B. Yaden, João Sedoc, Robert~J. DeRubeis, Robb Willer, and Johannes~C. Eichstaedt. 2024.
\newblock \href {https://doi.org/10.1038/s44184-024-00056-} {Large language models could change the future of behavioral healthcare: a proposal for responsible development and evaluation}.
\newblock \emph{npj Mental Health Res}.

\bibitem[{Strachan et~al.(2024)Strachan, Albergo, Borghini, and et~al.}]{strachan2024testing}
J.~W.~A. Strachan, D.~Albergo, G.~Borghini, and et~al. 2024.
\newblock \href {https://doi.org/10.1038/s41562-024-01882-z} {Testing theory of mind in large language models and humans}.
\newblock \emph{Nature Human Behaviour}.

\bibitem[{Sun et~al.(2024)Sun, Huang, Wang, Wu, Zhang, Li, Gao, Huang, Lyu, Zhang, Li, Liu, Liu, Wang, Zhang, Vidgen, Kailkhura, Xiong, Xiao, Li, Xing, Huang, Liu, Ji, Wang, Zhang, Yao, Kellis, Zitnik, Jiang, Bansal, Zou, Pei, Liu, Gao, Han, Zhao, Tang, Wang, Vanschoren, Mitchell, Shu, Xu, Chang, He, Huang, Backes, Gong, Yu, Chen, Gu, Xu, Ying, Ji, Jana, Chen, Liu, Zhou, Wang, Li, Zhang, Wang, Xie, Chen, Wang, Liu, Ye, Cao, Chen, and Zhao}]{sun2024trustllm}
Lichao Sun, Yue Huang, Haoran Wang, Siyuan Wu, Qihui Zhang, Yuan Li, Chujie Gao, Yixin Huang, Wenhan Lyu, Yixuan Zhang, Xiner Li, Zhengliang Liu, Yixin Liu, Yijue Wang, Zhikun Zhang, Bertie Vidgen, Bhavya Kailkhura, Caiming Xiong, Chaowei Xiao, Chunyuan Li, Eric Xing, Furong Huang, Hao Liu, Heng Ji, Hongyi Wang, Huan Zhang, Huaxiu Yao, Manolis Kellis, Marinka Zitnik, Meng Jiang, Mohit Bansal, James Zou, Jian Pei, Jian Liu, Jianfeng Gao, Jiawei Han, Jieyu Zhao, Jiliang Tang, Jindong Wang, Joaquin Vanschoren, John Mitchell, Kai Shu, Kaidi Xu, Kai-Wei Chang, Lifang He, Lifu Huang, Michael Backes, Neil~Zhenqiang Gong, Philip~S. Yu, Pin-Yu Chen, Quanquan Gu, Ran Xu, Rex Ying, Shuiwang Ji, Suman Jana, Tianlong Chen, Tianming Liu, Tianyi Zhou, William Wang, Xiang Li, Xiangliang Zhang, Xiao Wang, Xing Xie, Xun Chen, Xuyu Wang, Yan Liu, Yanfang Ye, Yinzhi Cao, Yong Chen, and Yue Zhao. 2024.
\newblock \href {https://arxiv.org/abs/2401.05561} {Trustllm: Trustworthiness in large language models}.
\newblock \emph{Proceedings of the 41st International Conference on Machine Learning (ICML)}, arXiv:2401.05561.

\bibitem[{Tan et~al.(2024)Tan, Elangovan, Ong, Shah, Sung, Wong, Xue, Liu, Wang, Kuo, Chesterman, Yeong, and Ting}]{tan2024proposedscoreevaluationframework}
Ting~Fang Tan, Kabilan Elangovan, Jasmine Ong, Nigam Shah, Joseph Sung, Tien~Yin Wong, Lan Xue, Nan Liu, Haibo Wang, Chang~Fu Kuo, Simon Chesterman, Zee~Kin Yeong, and Daniel~SW Ting. 2024.
\newblock \href {https://arxiv.org/abs/2407.07666} {A proposed s.c.o.r.e. evaluation framework for large language models : Safety, consensus, objectivity, reproducibility and explainability}.
\newblock \emph{Preprint}, arXiv:2407.07666.

\bibitem[{Templeton et~al.(2024)Templeton, Conerly, Marcus, Lindsey, Bricken, Chen, Pearce, Citro, Ameisen, Jones, Cunningham, Turner, McDougall, MacDiarmid, Freeman, Sumers, Rees, Batson, Jermyn, Carter, Olah, and Henighan}]{templeton2024scaling}
A.~Templeton, T.~Conerly, J.~Marcus, J.~Lindsey, T.~Bricken, B.~Chen, A.~Pearce, C.~Citro, E.~Ameisen, A.~Jones, H.~Cunningham, N.~L. Turner, C.~McDougall, M.~MacDiarmid, C.~D. Freeman, T.~R. Sumers, E.~Rees, J.~Batson, A.~Jermyn, S.~Carter, C.~Olah, and T.~Henighan. 2024.
\newblock \href {https://transformer-circuits.pub/2024/scalingmonosemanticity/index.html} {Scaling monosemanticity: Extracting interpretable features from claude 3 sonnet}.
\newblock In \emph{Transformer Circuits Thread}.
\newblock [Online].

\bibitem[{Thirunavukarasu et~al.(2023)Thirunavukarasu, Ting, Elangovan, and et~al.}]{thirunavukarasu2023}
A.J. Thirunavukarasu, D.S.J. Ting, K.~Elangovan, and et~al. 2023.
\newblock \href {https://doi.org/10.1038/s41591-023-02448-8} {Large language models in medicine}.
\newblock \emph{Nat Med}, 29:1930--1940.

\bibitem[{Vats et~al.(2023)Vats, Zope, De, Sharma, Bhattacharya, Nigam, Guha, Rudra, and Ghosh}]{vats2023llms}
Shaurya Vats, Atharva Zope, Somsubhra De, Anurag Sharma, Upal Bhattacharya, Shubham~Kumar Nigam, Shouvik~Kumar Guha, Koustav Rudra, and Kripabandhu Ghosh. 2023.
\newblock \href {https://openreview.net/forum?id=DgNnVebNPy} {{LLM}s -- the good, the bad or the indispensable?: A use case on legal statute prediction and legal judgment prediction on indian court cases}.
\newblock In \emph{The 2023 Conference on Empirical Methods in Natural Language Processing}.

\bibitem[{Wang and Zhou(2024)}]{CoTRWOP}
Xuezhi Wang and Denny Zhou. 2024.
\newblock \href {https://arxiv.org/abs/2402.10200} {Chain-of-thought reasoning without prompting}.
\newblock \emph{Preprint}, arXiv:2402.10200.

\bibitem[{Wang et~al.(2023)Wang, Zhao, and Petzold}]{wang2023large}
Yuqing Wang, Yun Zhao, and Linda Petzold. 2023.
\newblock \href {https://arxiv.org/abs/2304.05368} {Are large language models ready for healthcare? a comparative study on clinical language understanding}.
\newblock \emph{Preprint}, arXiv:2304.05368.

\bibitem[{Wei et~al.(2022)Wei, Wang, Schuurmans, Bosma, Ichter, Xia, Chi, Le, and Zhou}]{wei2023chainofthought}
Jason Wei, Xuezhi Wang, Dale Schuurmans, Maarten Bosma, Brian Ichter, Fei Xia, Ed~Chi, Quoc Le, and Denny Zhou. 2022.
\newblock \href {https://arxiv.org/abs/2201.11903} {Chain-of-thought prompting elicits reasoning in large language models}.
\newblock \emph{36th Conference on Neural Information Processing Systems (NeurIPS 2022)}, arXiv:2201.11903.

\bibitem[{Wu et~al.(2023)Wu, Irsoy, Lu, Dabravolski, Dredze, Gehrmann, Kambadur, Rosenberg, and Mann}]{wu2023bloomberggpt}
Shijie Wu, Ozan Irsoy, Steven Lu, Vadim Dabravolski, Mark Dredze, Sebastian Gehrmann, Prabhanjan Kambadur, David Rosenberg, and Gideon Mann. 2023.
\newblock \href {https://arxiv.org/abs/2303.17564} {Bloomberggpt: A large language model for finance}.
\newblock \emph{Preprint}, arXiv:2303.17564.

\bibitem[{Xiong et~al.(2024)Xiong, Hu, Lu, Li, Fu, He, and Hooi}]{xiong2024llms}
Miao Xiong, Zhiyuan Hu, Xinyang Lu, Yifei Li, Jie Fu, Junxian He, and Bryan Hooi. 2024.
\newblock \href {https://arxiv.org/abs/2306.13063} {Can llms express their uncertainty? an empirical evaluation of confidence elicitation in llms}.
\newblock \emph{Preprint}, arXiv:2306.13063.

\bibitem[{Xu and Cohen(2018)}]{xu-cohen-2018-stock}
Yumo Xu and Shay~B. Cohen. 2018.
\newblock \href {https://doi.org/10.18653/v1/P18-1183} {Stock movement prediction from tweets and historical prices}.
\newblock In \emph{Proceedings of the 56th Annual Meeting of the Association for Computational Linguistics (Volume 1: Long Papers)}, pages 1970--1979, Melbourne, Australia. Association for Computational Linguistics.

\bibitem[{Yadkori et~al.(2024)Yadkori, Kuzborskij, György, and Szepesvári}]{yadkori2024believe}
Yasin~Abbasi Yadkori, Ilja Kuzborskij, András György, and Csaba Szepesvári. 2024.
\newblock \href {https://arxiv.org/abs/2406.02543} {To believe or not to believe your llm}.
\newblock \emph{Preprint}, arXiv:2406.02543.

\bibitem[{Yang et~al.(2023)Yang, Tang, and Tam}]{yang2023investlm}
Yi~Yang, Yixuan Tang, and Kar~Yan Tam. 2023.
\newblock \href {https://arxiv.org/abs/2309.13064} {Investlm: A large language model for investment using financial domain instruction tuning}.
\newblock \emph{Preprint}, arXiv:2309.13064.

\bibitem[{Zhang et~al.(2023{\natexlab{a}})Zhang, Yang, and Liu}]{zhang2023instructfingpt}
Boyu Zhang, Hongyang Yang, and Xiao-Yang Liu. 2023{\natexlab{a}}.
\newblock \href {https://arxiv.org/abs/2306.12659} {Instruct-fingpt: Financial sentiment analysis by instruction tuning of general-purpose large language models}.
\newblock \emph{Preprint}, arXiv:2306.12659.

\bibitem[{Zhang et~al.(2023{\natexlab{b}})Zhang, Yang, Zhou, Ali~Babar, and Liu}]{zhang2023enhancing}
Boyu Zhang, Hongyang Yang, Tianyu Zhou, Muhammad Ali~Babar, and Xiao-Yang Liu. 2023{\natexlab{b}}.
\newblock \href {https://arxiv.org/pdf/2310.04027} {Enhancing financial sentiment analysis via retrieval augmented large language models}.
\newblock In \emph{Proceedings of the Fourth ACM International Conference on AI in Finance}, pages 349--356.

\bibitem[{Zhao et~al.(2024{\natexlab{a}})Zhao, Chen, Yang, Liu, Deng, Cai, Wang, Yin, and Du}]{10.1145/3639372}
Haiyan Zhao, Hanjie Chen, Fan Yang, Ninghao Liu, Huiqi Deng, Hengyi Cai, Shuaiqiang Wang, Dawei Yin, and Mengnan Du. 2024{\natexlab{a}}.
\newblock \href {https://doi.org/10.1145/3639372} {Explainability for large language models: A survey}.
\newblock \emph{ACM Trans. Intell. Syst. Technol.}, 15(2).

\bibitem[{Zhao et~al.(2024{\natexlab{b}})Zhao, Yan, Sun, Xing, Wang, Meng, Cheng, Ren, and Yin}]{zhao-etal-2024-improving}
Yukun Zhao, Lingyong Yan, Weiwei Sun, Guoliang Xing, Shuaiqiang Wang, Chong Meng, Zhicong Cheng, Zhaochun Ren, and Dawei Yin. 2024{\natexlab{b}}.
\newblock \href {https://aclanthology.org/2024.lrec-main.782} {Improving the robustness of large language models via consistency alignment}.
\newblock In \emph{Proceedings of the 2024 Joint International Conference on Computational Linguistics, Language Resources and Evaluation (LREC-COLING 2024)}, pages 8931--8941, Torino, Italia. ELRA and ICCL.

\bibitem[{Zheng et~al.(2024)Zheng, Zhu, Lin, Lu, Han, and Sun}]{zheng2024executing}
Xin Zheng, Qiming Zhu, Hongyu Lin, Yaojie Lu, Xianpei Han, and Le~Sun. 2024.
\newblock \href {https://arxiv.org/abs/2403.00795} {Executing natural language-described algorithms with large language models: An investigation}.
\newblock \emph{Preprint}, arXiv:2403.00795.

\end{thebibliography}
\newpage

\appendix

\section{Related Work} \label{app:literature}

A brief summary of recent research in mission-critical domains of law, health, and finance, in the context of LLMs, has been presented here.

In the medical domain, LLMs are recognized for their ability to incorporate clinical knowledge, as highlighted by \cite{Rae2023}. Consequently, there is significant interest in the proposal of utilizing LLMs for medical applications \cite{chen2024evaluating}, for instance, understanding clinical language \cite{wang2023large}. Moreover, the utility of LLMs in clinical research has been underscored \cite{Cascella2023EvaluatingTF}.  In spite of this, as noted by \cite{Karabacak2023embracing} and \cite{deangelis2023chatgpt}, there are significant challenges associated with utilizing LLMs in medical domain.

Within the financial domain, notable advancements include BloombergGPT, the first financial LLM trained from scratch \cite{wu2023bloomberggpt}. On a similar note, \cite{yang2023investlm}, \cite{chen2023discfinllm}, deal with fine-tuning LLMs specifically for financial tasks, such as investing. Considerable recent work has focused on exploring the utility of LLMs in performing financial tasks, including financial sentiment analysis \cite{zhang2023instructfingpt}. Interestingly, \cite{kim2024financial} has concluded that prediction accuracy of LLM is at par with state-of-the-art Machine learning (ML) model.
Still, \cite{lakkaraju2023llms} has identified critical gaps in providing reliable financial information using LLM. 

In the realm of law, \cite{katz2023} has established that GPT-4 by OpenAI has enough legal knowledge and understanding to pass the bar exam. As a result, LLMs have been used successfully to perform many tasks \cite{lai2023large}, including summarizing the judgment of legal cases \cite{ghosh2023}.

\section{Trustworthiness in LLMs}
\label{app:CoT}

{

\cite{sun2024trustllm} ranks trustworthiness of 16 LLMs on truthfulness, safety, fairness, privacy, machine ethics and robustness. In this paper, we are primarily concerned with the aspect of robustness. A major contributing factor towards a more robust LLM is the reproducibility of its output. Reproducibility is an important aspect of trustworthiness, particularly in the context of high-stakes applications. This research paper focuses on three mission-critical domains: law, health, and finance -- where AI without reproducibility would not be acceptable by domain experts.

Despite the challenges in ensuring the reproducibility of LLM outputs and the vast wealth of literature on the trustworthiness of LLM, there remain notable gaps in reproducibility studies within specific domains such as finance and law. While \cite{Franc2024} and \cite{Most2024.03.21.24304667} have explored the reproducibility of LLM performance in emergency department triage and pharmacy education, respectively, our study utilizes distinct datasets and introduces a novel technique.
}

\section{Reproducibility metric}
\label{app:reproducibility}
{
Although the reproducibility metric defined in section \ref{sec:rep}, is broadly applicable due to its inherent versatility, in this paper, its application is confined to high-stakes domains where the outcomes are pivotal given their direct consequences on human well-being. For example in the domain of law, statute prediction is an important problem \cite{10.1145/3626772.3657879}, that can aid in law enforcement and in determining the nature and severity of punishment. Similarly, predicting the future behavior of stocks is critical for market investors in finance. In the same vein, predicting diseases and corresponding medicines can be extremely useful in health sector, especially in rural healthcare systems where doctors are not always available. Such AI based systems can assist primary-level treatment, as vouched by practicing doctors at a medical institute of national importance. \cite{Khan2021} highlights the acute shortage of doctors in developing countries with large populations like India and Indonesia, resulting in a significant deviation from the WHO norm of at least one doctor per 1,000 patients. In short, we chose to apply the reproducibility metric to three critical domains which directly impact human lives.

For completeness, we provide a brief discussion of other robustness metrics available in the literature. In \cite{zhao-etal-2024-improving} robustness for Large Language Models (LLMs) centers on consistency of response generation, which is ultimately determined by using an LLM (GPT-4) for judging whether the response was consistent. In the ReQuest framework LLM is not used for judging as their judgment may not always be reliable and moreover our metric employs strict character to character match. \cite{Franc2024} measures the variability in output caused by variations in the prompt wording. However, the ReQuest method does not control for prompt wording since, in real-world applications, users may create queries beyond a specific template. 

In \cite{kim2024generativeartificialintelligencereproducibility}, perceptual hashes are utilized to capture content essence instead of comparing raw outputs (e.g., images or text), differing from the exact comparisons crucial in high-stakes domains, as emphasized in our work. \cite{sarker2024syntacticrobustnessllmbasedcode} has designed a simple robustness metric for an LLM-based code generator that has been considered syntactically robust if it produces semantically equivalent code. The S.C.O.R.E. metric proposed in \cite{tan2024proposedscoreevaluationframework} focusses on reproducibility in terms of semantic consistency rather than strict word-for-word match. In contrast, our more rigorous approach to reproducibility ensures greater suitability for high-stakes domains and is broadly applicable across various fields, while avoiding reliance on LLM judgment.

}

\section{Results summary for Human Rights violation}\label{results:HR}
{
For \texttt{gemini} (Intra-LLM setup), PerRR\_GP\_G and PreRR\_GP\_G are observed to be 92.011 and 0.9613, respectively, while for \texttt{llama3} (Intra-LLM setup), these values are recorded as 95.859 and 0.9609.

 While the Intra-LLM metrics for \texttt{gemini} align closely with those presented in Table \ref{tab:legal}, the \texttt{llama3} configuration demonstrates superior performance in the Intra-LLM setup across the entire dataset comprising 2,377 cases. In the Inter-LLM setup, specifically for the \texttt{gemini} $\rightarrow$ \texttt{llama3} configuration, the PerRR\_GP\_L, PreRR\_GP\_L, PerRR\_GA\_L, and PreRR\_GA\_L values are 93.453, 0.9377, 98.433, and 0.9579, respectively.

 Conversely, the corresponding values for the \texttt{llama3} $\rightarrow$ \texttt{gemini} setup, specifically PerRR\_LP\_G, PreRR\_LP\_G, PerRR\_LA\_G, and PreRR\_LA\_G, are 95.396, 0.9550, 91.602, and 0.9327, respectively. Notably, for the larger dataset, the results are either comparable or demonstrate improvement, particularly in the case of PerRR\_LA\_G for the \texttt{llama3} $\rightarrow$ \texttt{gemini} configuration.

To evaluate LLM agnosticism, we assess whether algorithms generated by one LLM generalize to another. \cite{zheng2024executing} shows that state-of-the-art LLMs can effectively execute natural language-described algorithms. By testing the same algorithm across two LLMs, we can measure their procedural consistency.

 }

\section{Results for financial tasks} \label{app:finance}
In the intra-LLM setup for Apple (AAPL), we observed the performance of \texttt{gemini} and \texttt{llama3-70b} models using ReQuest algorithm $\mathcal{A}$. The baseline Macro F1 score for BERTweet was 0.53. When \texttt{gemini} was employed, the initial Macro F1 score was 0.44, which improved to 0.48 after applying $\mathcal{A}$. This improvement in accuracy was reflected in the high reproducibility percentage (PerRR\_GP\_G) of 91\%. On the other hand, \texttt{llama3-70b} started with a Macro F1 score of 0.41 and showed a slight decrease to 0.38 with ($\mathcal{A}$) resulting in a PerRR\_LP\_L of 92.68\%, indicating high reproducibility but a drop in performance. In the Inter-LLM setup, we examined the cross-model reproducibility where ReQuest algorithm $\mathcal{A}$ generated by \texttt{gemini} was applied to \texttt{llama3-70b} and vice versa. When ReQuest algorithm $\mathcal{A}$ from \texttt{gemini} was applied to \texttt{llama3-70b}, the PerRR\_GP\_L was 91.67\%, suggesting a high degree of reproducibility, though the Macro F1 score was moderately lower at 0.48. Conversely, applying ReQuest algorithm $\mathcal{A}$ from \texttt{llama3-70b} to \texttt{gemini} resulted in a PerRR\_LP\_G of 78.84\%, indicating a lower reproducibility and a Macro F1 score of 0.52.

The evaluations for Google (GOOG) followed a similar pattern. The baseline Macro F1 score for BERTweet was 0.55. For \texttt{gemini}, the initial Macro F1 score was 0.45, which slightly improved to 0.46 after applying ReQuest algorithm $\mathcal{A}$ with a PerRR\_GP\_G of 97.82\%, showing high reproducibility. \texttt{llama3}'s initial Macro F1 score of 0.37 decreased to 0.35 with $\mathcal{A}$, achieving a PerRR\_LP\_L of 94.59\%. In the Inter-LLM setup for Google, the cross-model reproducibility was again assessed. When ReQuest algorithm $\mathcal{A}$ from \texttt{gemini} was applied to \texttt{llama3}, the PerRR\_GP\_L was 93.75\%, and the Macro F1 score was 0.48. Applying $\mathcal{A}$ from \texttt{llama3} to \texttt{gemini} resulted in a PerRR\_LP\_G of 71.15\%, with a lower Macro F1 score of 0.52.

{
Results for two new datasets have been added in Table \ref{tab:financeadditional}, where the stock movement for Microsoft and Amazon were predicted. Similar trends were observed for the Microsoft (MSFT) and Amazon (AMZN) datasets. For the Microsoft dataset, the baseline Macro F1 score with BERTweet was 0.51. \texttt{Gemini}'s performance improved from 0.42 to 0.51 with ReQuest, achieving a PerRR\_GP\_G of 82\%. Upon employing \texttt{llama3}, the Macro F1 score declined from 0.41 to 0.37, despite a high PerRR\_LP\_L of 90.2\%. In the inter-LLM setup, \texttt{gemini}'s ReQuest applied to \texttt{llama3} achieved a high PerRR\_GP\_L of 95.45\%, while \texttt{llama3} to \texttt{gemini} showed a PerRR\_LP\_G of 91.11\%. For the Amazon dataset, \texttt{gemini} exhibited an increase in performance with ReQuest, improving from 0.47 to 0.49, whereas \texttt{llama3} demonstrated a decline from 0.43 to 0.37. The inter-LLM reproducibility for the Amazon dataset was observed to be lower than that for the Microsoft dataset, with PerRR\_GP\_L recorded at 82.97\% and PerRR\_LP\_G at 87.75\%. 
}

\section{Results Summary for health tasks} \label{app:health}
In $T_1$, the {\bf{PreRR\_GP\_G}} score of 0.617 indicates that the algorithm generated by \texttt{gemini-1.0-pro} is able to replicate a decent proportion of its initial predictions at the data point level, as is the case with \texttt{llama3} with a {\bf{PreRR\_LP\_L}} score of 0.701. But for $T_2$, low scores of both {\bf{PreRR\_GP\_G}} and {\bf{PreRR\_LP\_L}} indicate the reasoning process of LLMs may be inconsistent and highly sensitive to the prompt structure and wording. 

For $T_1$, in the ``inter-LLM'' setup, when \texttt{gemini}'s algorithm, $\mathcal{A}$ is executed on \texttt{llama3} (\texttt{gemini} $\rightarrow$ \texttt{llama3}), the \textbf{PerRR\_GP\_L} score of 95.47\% conveys very consistent performance when compared to \texttt{gemini}'s task prompt responses, which suggests the reasoning processes of the two LLMs are in parity. Further, the green arrow reveals \texttt{llama3}'s responses using \texttt{gemini's} $\mathcal{A}$ are better than \texttt{gemini}'s task prompt responses. A \textbf{PreRR\_GP\_L} score of 0.728 indicates a healthy number of exact matches, which, though ever so slightly, increases to 0.731 when \texttt{gemini}'s $\mathcal{A}$ is `executed' on both \texttt{gemini} and \texttt{llama3} (\textbf{PreRR\_GA\_L}). A low score of (\textbf{PerRR\_GA\_L}) 86.47\% indicates \texttt{llama3}'s reasoning is not analogous to \texttt{gemini}'s $\mathcal{A}$ even though the former responds better with \texttt{gemini}'s $\mathcal{A}$. 
In case of \texttt{llama3} $\rightarrow$ \texttt{gemini}, a near perfect score of \textbf{PerRR\_LP\_G} (99.31\%) may indicate a striking similarity in \texttt{llama3}'s internal mechanism and it's $\mathcal{A}$. Further, the high value of \textbf{PerRR\_LA\_G} (94.99\%) accounts for the overlap between results of \texttt{llama3}'s algorithm, $\mathcal{A}$ being executed on \texttt{llama3} and \texttt{gemini}. However, this does not take away the fact that their predictions are not in correspondence with the true labels in the dataset. 

For $T_2$, while \texttt{gemini}'s \textbf{PerRR\_GP\_G} is higher (97.545\%), the \textbf{PreRR\_GP\_G} of 0.672 is still low, suggesting that individual predictions are not reproducible in totality.
\texttt{llama3}, however, shows a significant incredible performance in both \textbf{PerRR\_LP\_L} and \textbf{PreRR\_LP\_L}, implying that its algorithms are very effective in replicating its original performance, especially at the individual prediction level. 
\newline
Unlike $T_1$, $\mathcal{A}$ produced by both \texttt{gemini} and \texttt{llama3} performs poorly for the inter-LLM setup in $T_2$.

\section{Why use PreRR when we already have Macro-F1 based PerRR?}\label{sec:PreRRvsPerRR}
A scenario in which the PerRR score may be misleading is when the macro f1 score of any 2 prediction sets to be evaluated is low especially for a relatively simple task like binary classification where the odds of being right are 50\% for each data point. In such cases there might be more than one unique prediction set to yield the same low macro-f1 score. This will result in multiple unique and pred\_sets pairs where the PerRR is 100\%. An example of such case is given below:\\
Suppose we have a binary classification problem on a dataset of size 6. The gold standard labels are given by {\it gold\_std} = \{1,0,1,0,1,0\} i.e it is a perfectly balanced dataset. Lets say LLM1 and LLM2 predicted {\it pred\_set\_1} = \{0,0,0,0,1,1\} and {\it pred\_set\_2} = \{0,0,0,1,1,0\} respectively. So Macro\_F1({\it pred\_set\_1}) = Macro\_F1({\it pred\_set\_2}) = 0.4857. Hence PerRR({\it pred\_set\_1},{\it pred\_set\_2}) = 100\%. But {\it pred\_set\_1} and {\it pred\_set\_2} are clearly different, leading to a misleading sense of reproducibility if one was to only rely on PerRR. In fact in this particular example there are 18 such {\it pred\_sets} which will have 100\% PerRR with {\it pred\_set\_1}. This is actually the peak of the distribution given in Figure {\ref{fig:balanced_binary}}. This calls for necessity to introduce a stricter measure of similarity for such cases, which is fulfilled by the PreRR metric. Here the PreRR score will be 0.66 which accurately depicts the difference between both pred\_sets.\\ 
If the complexity of the task is increased to a 3-class multi-class classification problem on same size dataset, most of the distribution shifts towards left as shown in Figure \ref{fig:multi_class} where the {\it gold\_std} = \{0,1,2,0,1,2\}. Following the same behaviour for even more complex task like multi-label classification, the distribution should shift even more towards the left. This distribution directly signifies the probability of discrepancy between the PerRR and PreRR score for a given classification problem. As most of the distribution is located in the left half of the Macro-F1 axis, the robustness of PerRR score decreases for low macro-F1 score for simple problems like binary classification and very low macro-F1 score for more complex tasks like multi-label classification.\\
Due to inherent strict nature of PreRR, its variance should increase as the task complexity increases. Simpler tasks like binary classification should have little variations between pred\_sets. For more complex tasks like multi-label classification, it should vary more relative to PerRR for slightest of mistakes (especially at lower macro-F1s). 

\section{Future Work}

Given the results on reproducibility, it would be interesting to revisit the criteria proposed by \cite{herrmann2024standards} for belief representation in LLMs — accuracy, coherence, uniformity, and use. 

\cite{strachan2024testing} has conducted tests to evaluate LLMs on theory of mind tasks. They have found that across a battery of tasks, GPT-4 has performed at human level. Nevertheless, based on our findings we argue that LLMs simulate aspects of human-like theory of mind rather than posses genuine theory of mind capabilities. Because, if LLMs genuinely possessed theory of mind capabilities reproducibility and consistency of model responses would be higher. Consequently, to be certain that GPT-4 is simulating theory of mind rather than possess genuine capacity for it, a natural followup to this study would be to test the reproducibility of GPT-4 responses, which could not be covered in this work due to financial costs involved.

\centering
\begin{table*}[]
    \centering

    \begin{tabular}{|p{16cm}|}
    \hline
         Definition of some variables
\{training\_prompt\}:\\

Statute ID: Indian Penal Code, 1860\_147
Title: Punishment for rioting
Description: Whoever is guilty of rioting, shall be punished with imprisonment of either description for a term which may extend to two years, or with fine, or with both.\\
\#\#\#\\
Statute ID: Indian Penal Code, 1860\_149\\
Title: Every member of unlawful assembly guilty of offence committed in prosecution of common object\\
Description: If an offence is committed by any member of an unlawful assembly in prosecution of the common object of mat assembly, or such as the members of that assembly knew to be likely to be committed in prosecution of that object, every person who, at the time of the committing of that offence, is a member of the same assembly, is guilty of that offence.\\
............(for 18 statutes)\\
\#\#\#\\
\hline
Task:\\
  You are given a fact statement delimited by triple backticks and statutes with
  their title and description. Your task is to identify the statutes applicable to the
  fact statement from the given statutes that you are most confident apply to the fact
  statement. Each statute consists of a title and a description of its scope and provisions.
  Include only those statute in your response which description logically matches with
  some parts of the fact statement.
  Training:\\
  Statutes:\\
  \{training\_prompt\}\\
  Fact Statement:\\
  ```\{inp\}'''\\
  Response and Instructions:\\
  Format of response: Statute1; Statute2 ...\\
  Your response should include the statutes applicable to the fact statement. The
  applicable statute must be mentioned exactly as it appears in Statutes provided.
  Include only those statutes which you are very sure about.\\
\hline
       
    \end{tabular}
    \caption{Task Prompt: Statute Prediction (Gemini); inspired by \cite{vats2023llms}}
    \label{tab:statute_task_prompt_gemini}
\end{table*}

\begin{table*}[]
    \centering

    \begin{tabular}{|p{16cm}|}

    \hline

Task:\\
You are given a fact statement delimited by triple backticks and statutes with \
their title and description. Your task is to identify the statutes applicable to the \
fact statement from the given statutes that you are most confident apply to the fact \
statement. Each statute consists of a title and a description of its scope and provisions. \
Include only those statute in your response which description logically matches with \
some parts of the fact statement.\\
Training:\\
Statutes:\\
\{training\_prompt\}\\
Fact Statement:\\
```\{inp\}```\\

Your response should include the statutes applicable to the fact statement. The \
applicable statute must be mentioned exactly as it appears in Statutes provided. \
Include only those statutes which you are very sure about. No need to provide any explanation. Return output in the format of response. \
No need to write 'Here is the response: '\\
Format of response: Statute1; Statute2 …
\\
\hline
    \end{tabular}
    \caption{Task Prompt: Statute Prediction (Llama3-70b); inspired by \cite{vats2023llms}}
    \label{tab:statute_task_prompt_Llama}
\end{table*}

\begin{table*}[]
    \centering

    \begin{tabular}{|p{16cm}|}
    \hline
        1. **Read and understand the fact statement.** Identify the key legal issues in the fact statement.\\
2. **Identify the relevant area of law.** For example, if the fact statement involves a criminal offense, then the relevant area of law is criminal law.\\
3. **Research the relevant statutes.** This may involve using a legal database or consulting with a legal professional.\\
4. **Analyze the statutes and determine which ones are applicable to the fact statement.** Consider the following factors:\\
    * The elements of the offense\\
    * The defenses to the offense\\
    * The penalties for the offense\\
5. **Determine the strength of the evidence.** Consider the following factors:\\
    * The strength of the prosecution's case\\
    * The strength of the defense's case\\
    * The likelihood of a conviction\\

**Example:**\\

**Fact statement:** A person is charged with murder.\\

**Steps to identify applicable statutes:**\\

1. **Read and understand the fact statement.** The key legal issue in the fact statement is murder.\\
2. **Identify the relevant area of law.** The relevant area of law is criminal law.\\
3. **Research the relevant statutes.** The relevant statute is the Indian Penal Code, 1860\_302, which defines murder as ``whoever commits murder shall be punished with death, or imprisonment for life, and shall also be liable to fine.''\\
4. **Analyze the statute and determine if it is applicable to the fact statement.** The statute is applicable to the fact statement because the person is charged with murder.\\
5. **Determine the strength of the evidence.** This step is not necessary for the purpose of identifying applicable statutes. However, it may be relevant for other purposes, such as determining the likelihood of a conviction.\\
\hline
    \end{tabular}
    \caption{ReQuest Algorithm: Statute Prediction (Gemini)}
    \label{tab:statute_algo_gemini}
\end{table*}

\begin{table*}[]
    \centering

    \begin{tabular}{|p{16cm}|}
    \hline
**Step 1: Parse the Fact Statement**\\

* Break down the fact statement into individual events, actions, and circumstances.\\
* Identify the key entities involved, such as people, objects, and locations.\\

**Step 2: Extract Relevant Information**\\

* Extract relevant information from the fact statement, including:\\
	+ Actions taken by individuals (e.g., shooting, assaulting, surrendering)\\
	+ Consequences of actions (e.g., injury, death)\\
	+ Objects involved (e.g., weapons, vehicles)\\
	+ Locations and settings (e.g., village, police station)\\

**Step 3: Match with Statute Descriptions**\\

* Compare the extracted information with the descriptions of the statutes provided.\\
* Look for statutes that mention similar actions, consequences, objects, or settings.\\

**Step 4: Filter and Rank Statutes**\\

* Filter out statutes that do not match the extracted information.\\
* Rank the remaining statutes based on their relevance and similarity to the fact statement.\\

**Step 5: Select Applicable Statutes**\\

* Select the top-ranked statutes that are most relevant to the fact statement.\\
* Ensure that the selected statutes are applicable to the events, actions, and circumstances described in the fact statement.\\

**Step 6: Return the Response**\\

* Return the applicable statutes in the required format, separated by semicolons.\\

\hline
    \end{tabular}
    \caption{ReQuest Algorithm: Statute Prediction (Llama3-70b)}
    \label{tab:statute_algo_Llama}
\end{table*}

\begin{table*}[]
    \centering

    \begin{tabular}{|p{16cm}|}
    \hline
      Task:\\
You are a bot who strictly follows steps. You strictly follow algorithms with very deterministic output. You are given a fact statement delimited by triple backticks (```) and statutes with
their title and description. Your task is to follow the given steps to find out the statutes applicable. Each statute consists of a title and a description of its scope and provisions. \\

Statutes:\\
\{training\_prompt\}\\
\#\#\\
Steps to follow:\\
\{algo\}\\
\#\#\\
Fact Statement:\\
```\{inp\}```
Response and Instructions:\\
Format of response: Statute1; Statute2 ...\\

The applicable statutes must be mentioned exactly as it appears in Statutes provided.\\
Include only those statutes which you get with very high confidence from the algorithm. STRICTLY USE THE SPECIFIED STEPS FOR PREDICTION. DO NOT USE ANY OTHER ALGORITHM.\\
\hline
    \end{tabular}
    \caption{Robustness Check Prompt: Statute Prediction (Gemini). The algorithm  is in Table \ref{tab:statute_algo_gemini}.}
    \label{tab:robustness_statute_gemini}
\end{table*}

\begin{table*}[]
    \centering

    \begin{tabular}{|p{16cm}|}
    \hline
      Task:\\
You are a bot who strictly follow steps. You strictly follow algorithms with very deterministic output. You are given a fact statement delimited by triple backticks (“‘) and statutes with
their title and description. Your task is to follow the given steps to find out the statutes applicable. Each statute consists of a title and a description of its scope and provisions.

Statutes:\\
\{training\_prompt\}\\
\#\#\\
Steps to follow:\\
\{algo\}\\
\#\#\\
Fact Statement:\\
```\{inp\}```\\
What relevant statutes did you get after following these steps for this fact statement? Return them in the following response format:\\
Format of response: Statute1; Statute2 ...\\
Your response should be such that I can extract the statutes using str.split(';'). Do not return anything else.\\

\hline
    \end{tabular}
    \caption{Robustness Check Prompt: Statute Prediction (Llama3-70b). The algorithm  is in Table \ref{tab:statute_algo_Llama}.}
    \label{tab:robustness_statute_Llama}
\end{table*}

\begin{table*}[]
    \centering

    \begin{tabular}{|p{10cm}|}
    \hline
import re\\
\\
def identify\_applicable\_statutes(fact\_statement):\\

  \hspace{0.5cm} \# Identify the key facts in the fact statement.\\
  \hspace{0.5cm}key\_facts = []\\
  \hspace{0.5cm} for line in fact\_statement.split("\textbackslash n"):\\
    \hspace{1cm}line = line.strip()\\
    \hspace{1cm}if line:\\
      \hspace{1.5cm}key\_facts.append(line)\\

  \hspace{0.5cm}\# Determine which statutes may be applicable to the facts.\\
  \hspace{0.5cm}potentially\_applicable\_statutes = []\\
  \hspace{0.5cm}for statute in statutes:\\
    \hspace{1cm}for key\_fact in key\_facts:\\
      \hspace{1.5cm}if re.search(key\_fact, statute["description"]):\\
        \hspace{2cm}potentially\_applicable\_statutes.append(statute)\\

  \hspace{0.5cm}\# Analyze the statutes to determine if they are actually applicable to the facts.\\
  \hspace{0.5cm}applicable\_statutes = []\\
  \hspace{0.5cm}for statute in potentially\_applicable\_statutes:\\
    \hspace{1cm}if is\_statute\_applicable(fact\_statement, statute):\\
      \hspace{1.5cm}applicable\_statutes.append(statute)\\

  \hspace{0.5cm} \# Return the list of applicable statutes.\\
  \hspace{0.5cm}return applicable\_statutes\\
\\

def is\_statute\_applicable(fact\_statement, statute):\\

  \hspace{0.5cm}\# Check if the fact statement contains any of the key terms from the statute's description.\\
  \hspace{0.5cm}for key\_term in statute["description"].split(" "):\\
    \hspace{1cm}if re.search(key\_term, fact\_statement):\\
      \hspace{1.5cm}return True\\

  \hspace{0.5cm}\# Check if the fact statement describes a situation that is covered by the statute.\\
  \hspace{0.5cm}for example in statute["examples"]:\\
    \hspace{1cm}if re.search(example, fact\_statement):\\
      \hspace{1.5cm}return True\\

  \hspace{0.5cm}\# Otherwise, the statute is not applicable.\\
  \hspace{0.5cm}return False\\
\hline
    \end{tabular}
    \caption{Python code generated by Gemini for Statute Prediction Task}
    \label{tab:statute_python_gemini}
\end{table*}

\begin{table*}[]
    \centering

    \begin{tabular}{|p{10cm}|}
    \hline
import re\\
\\
def identify\_applicable\_statutes(fact\_statement):\\

  \hspace{0.5cm}key\_facts = []\\
  \hspace{0.5cm}for line in fact\_statement.split("\textbackslash n"):\\
    \hspace{1cm}line = line.strip()\\
    \hspace{1cm}if line:\\
      \hspace{1.5cm}key\_facts.append(line)\\

  \hspace{0.5cm}potentially\_applicable\_statutes = []\\
  \hspace{0.5cm}for statute in statutes:\\
    \hspace{1cm}for key\_fact in key\_facts:\\
      \hspace{1.5cm}if re.search(key\_fact, statute["description"]):\\
        \hspace{2cm}potentially\_applicable\_statutes.append(statute)\\

  \hspace{0.5cm}applicable\_statutes = []\\
  \hspace{0.5cm}for statute in potentially\_applicable\_statutes:\\
    \hspace{1cm}if is\_statute\_applicable(fact\_statement, statute):\\
      \hspace{1.5cm}applicable\_statutes.append(statute)\\

  \hspace{0.5cm}return applicable\_statutes\\
\\

def is\_statute\_applicable(fact\_statement, statute):\\

  \hspace{0.5cm}for key\_term in statute["description"].split(" "):\\
    \hspace{1cm}if re.search(key\_term, fact\_statement):\\
      \hspace{1.5cm}return True\\

  \hspace{0.5cm}for example in statute["examples"]:\\
    \hspace{1cm}if re.search(example, fact\_statement):\\
      \hspace{1.5cm}return True\\

  \hspace{0.5cm}return False\\
\hline
    \end{tabular}
    \caption{Python code (comments removed) generated by Gemini for Statute Prediction Task}
    \label{tab:statute_python_nocomments_gemini}
\end{table*}


\begin{table*}[]
    \centering
    \begin{tabular}{c|c|c|c|c|}
    \cline{2-5}
         &  \multicolumn{2}{|c|}{w.r.t. Task Prompt} & \multicolumn{2}{c|}{w.r.t. ReQuest Algo}\\
         \hline
    \multicolumn{1}{|c|}{\bf Python Code} & {\bf Per\_RR} & {\bf Pre\_RR} & {\bf Per\_RR} & {\bf Pre\_RR} \\   
      \hline   
     \multicolumn{1}{|c|} {With comments} & 100 & 0.612 &  98.15 & 0.6643\\
     \hline
     \multicolumn{1}{|c|} {Without comments} & 97.42 & 0.6063 &  99.33 & 0.6094\\
     \hline
    \end{tabular}
    \caption{Results obtained after "executing" Python code generated by Gemini by prompting (Table \ref{tab:robustness_statute_gemini})}
    \label{tab:python_results}
\end{table*}

\begin{table*}[]
    \centering

    \begin{tabular}{|p{16cm}|}
    \hline
The PerRR score of both the versions of the Python script (with comments and without comments) prediction (by `executing' it within Gemini itself) with respect to task prompt (Table \ref{tab:statute_task_prompt_gemini}) is very high. This shows that the performance of both the versions of the Python codes are similar with respect to the ReQuest algorithm and the task prompt. The PreRR score of both the versions is higher than the PreRR score of ReQuest algorithm predictions with respect to task prompt predictions. The Macro-F1 score of the Python script (with comments) `executed' on Gemini by prompting is equal to the Macro-F1 score of the predictions generated by the task prompt on the same. However the relatively lower PreRR score suggests that the predictions were not exactly replicated by the Python script despite resulting in the same Macro-F1 score. It was also observed that removing the comments from the python script (which would have given additional information to the LLM in form of natural language) decreased the PreRR score both with respect to the task prompt and ReQuest algorithm.  
\\
\hline
       
    \end{tabular}
    \caption{Insights gained from Table \ref{tab:python_results}}
    \label{tab:python_insights}
\end{table*}

\begin{table*}[]
    \centering

    \begin{tabular}{|p{16cm}|}
    \hline
        You are given a court statement delimited by triple backticks. Your task is to identify whether any \\
human rights violation occurred by studying the court statement. Refer to the website link given below for \\
relevant Articles and Protocols. Use only those articles and protocols mentioned in the website. \\
Website link: {\url{http://www.hri.org/docs/ECHR50.html}}\\
Court statement: ```\{query\}'''\\
Response format: 1 if any one of given article or protocol is violated and 0 if none of them are violated.\\
If some articles or protocols are violated, mention them at the end with relevant sentences from the statement that led to the violation. \\
NOTE: THERE CAN BE MULTIPLE VIOLATIONS IN A SINGLE STATEMENT. ONLY RETURN THOSE ARTICLES WHICH YOU ARE MOST CONFIDENT ABOUT ELSE RETURN 0.
\\
\hline
       
    \end{tabular}
    \caption{Task Prompt: Human Rights Violation Prediction (Gemini)}
    \label{tab:HR_task_prompt_gemini}
\end{table*}

\begin{table*}[]
    \centering

    \begin{tabular}{|p{16cm}|}
    \hline
You are given a court statement delimited by triple backticks. Your task is to identify whether any \
human rights violation occured by studying the court statement. Refer to the ECHR website link given below for \
relevant Articles and Protocols. Use only those articles and protocols mentioned in the website.\\
Website link: \url{http://www.hri.org/docs/ECHR50.html}\\
Court statement: ```\{query\}```\\
If some articles or protocols according to ECHR are violated, mention them at the end with relevant sentences from the statement that led to the violation. \
NOTE: THERE CAN BE MULTIPLE VIOLATIONS OR NO VIOLATIONS IN A SINGLE STATEMENT. ONLY RETURN THOSE ARTICLES WHICH YOU ARE MOST CONFIDENT ABOUT ELSE RETURN 0.\\
Response format: 1 if any one of given article or protocol is violated and 0 if none of them are violated. \
The 1 (if any article or protocol is violated) or 0 (if no article or protocol is violated) should be on the first line followed by a newline character.\\
\hline
       
    \end{tabular}
    \caption{Task Prompt: Human Rights Violation Prediction (Llama3-70b)}
    \label{tab:HR_task_prompt_Llama}
\end{table*}

\begin{table*}[]
    \centering

    \begin{tabular}{|p{16cm}|}
    \hline
      1. **Identify the relevant facts in the statement:**\\
   - Who is the applicant?\\
   - What are the applicant's claims?\\
   - What are the facts of the case?\\
   - What was the outcome of the case?\\

2. **Identify the relevant articles in the human rights treaty:**\\
   - Which human rights treaty is applicable to the case?\\
   - Which articles of the treaty are relevant to the applicant's claims?\\

3. **Analyze the facts in light of the relevant articles:**\\
   - Do the facts of the case violate any of the articles of the treaty?\\
   - If so, which articles are violated?\\

**Example:**\\

**Court statement:**\\

> The applicant is a journalist who was arrested and detained for three days without charge. He was not allowed to contact a lawyer or his family during his detention.\\

**Steps to identify human rights violations:**\\

1. **Identify the relevant facts in the statement:**\\
   - The applicant is a journalist.\\
   - The applicant was arrested and detained for three days without charge.\\
   - The applicant was not allowed to contact a lawyer or his family during his detention.\\

2. **Identify the relevant articles in the human rights treaty:**\\
   - The Universal Declaration of Human Rights (UDHR) is applicable to the case.\\
   - Articles 9, 10, and 11 of the UDHR are relevant to the applicant's claims.\\

3. **Analyze the facts in light of the relevant articles:**\\
   - Article 9 of the UDHR protects the right to liberty and security of person. The applicant's detention without charge violated his right to liberty and security of person.\\
   - Article 10 of the UDHR protects the right to freedom of expression. The applicant's arrest and detention for his journalistic activities violated his right to freedom of expression.\\
   - Article 11 of the UDHR protects the right to be treated with humanity and respect for the inherent dignity of the human person. The applicant's treatment during his detention, including being denied access to a lawyer and his family, violated his right to be treated with humanity and respect for his inherent dignity.\\

\hline
    \end{tabular}
    \caption{ReQuest Algorithm: Human Rights Violation Prediction (Gemini)}
    \label{tab:HR_algo_gemini}
\end{table*}

\begin{table*}[]
    \centering

    \begin{tabular}{|p{16cm}|}
    \hline
**Step 1: Read and understand the court statement**\\

* Read the entire court statement to understand the events, allegations, and circumstances described.\\
* Take note of key events, dates, times, locations, and individuals involved.\\

**Step 2: Identify potential human rights issues**\\

* Look for keywords and phrases that may indicate human rights violations, such as:\\
	+ Physical harm or abuse\\
	+ Coercion or intimidation\\
	+ Discrimination or bias\\
	+ Unfair treatment or procedure\\
	+ Violations of privacy or dignity\\
* Identify specific incidents or events that may constitute human rights violations.\\

**Step 3: Consult relevant human rights frameworks**\\

* Refer to relevant human rights frameworks, such as the European Convention on Human Rights (ECHR), the Universal Declaration of Human Rights (UDHR), or other applicable national or international human rights instruments.\\
* Familiarize yourself with the articles, protocols, and provisions that protect human rights.\\

**Step 4: Match allegations with human rights frameworks**\\

* Compare the allegations and incidents identified in Step 2 with the relevant human rights frameworks consulted in Step 3.\\
* Determine whether the allegations, if true, would constitute a violation of specific human rights provisions.\\

**Step 5: Analyze evidence and context**\\

* Evaluate the evidence presented in the court statement, including witness statements, documents, and other evidence.\\
* Consider the context in which the alleged violations occurred, including the circumstances, motivations, and intentions of the individuals involved.\\

**Step 6: Determine human rights violations**\\

* Based on the analysis in Step 5, determine whether the allegations constitute a violation of human rights provisions.\\
* Identify the specific articles, protocols, or provisions that are violated, and describe the nature of the violation.\\

**Step 7: Document findings**\\

* Clearly document the findings, including the specific human rights violations identified, the relevant provisions violated, and the evidence supporting the conclusion.\\
* Provide a concise and objective summary of the human rights violations identified in the court statement.\\

\hline
    \end{tabular}
    \caption{ReQuest Algorithm: Human Rights Violation Prediction (Llama3-70b)}
    \label{tab:HR_algo_Llama}
\end{table*}

\begin{table*}[]
    \centering

    \begin{tabular}{|p{16cm}|}
    \hline
Task:\\
You are a bot who strictly follow steps. You strictly follow algorithms with very deterministic output. You are given a court statement delimited by triple backticks (```). Your task is to follow the given steps to find out if any human right article or protocol is violated. Refer to the website link given below for \\
relevant Articles and Protocols: ONLY USE THOSE PROTOCOLS AND ARTICLES MENTIONED IN THE WEBSITE BELOW.\\
Website link: \url{http://www.hri.org/docs/ECHR50.html}\\
Steps to follow:\\
\{algo\}\\
\#\#\\
Court statement: ```{query}'''\\
\#\#\\
Format of response: 1 if any one of given article or protocol is violated and 0 if none of them are violated. Anyting else at the end.\\
STRICTLY USE THE SPECIFIED STEPS FOR PREDICTION. DO NOT USE ANY OTHER ALGORITHM. If some articles or protocols are violated, mention them at the end with the relevant steps provided by me that led you to that conclusion.\\
NOTE: THERE CAN BE MULTIPLE VIOLATIONS IN A SINGLE STATEMENT. ONLY RETURN THOSE ARTICLES (from ECHR) WHICH YOU ARE MOST CONFIDENT ABOUT ELSE RETURN 0.
\\
\hline
    \end{tabular}
    \caption{Robustness Check Prompt: Human Rights Violation (Gemini). The algorithm is in Table \ref{tab:HR_algo_gemini}.}
    \label{tab:robustness_HR_gemini}
\end{table*}

\begin{table*}[]
    \centering
    \begin{tabular}{|p{16cm}|}
    \hline
Task:\\
You are a bot who strictly follow steps. You strictly follow algorithms with very deterministic output. You are given a court statement delimited by triple backticks (“‘). \
Your task is to follow the given steps to find out if any articles or protocols according to ECHR are violated. Refer to the ECHR website link given below for \
relevant Articles and Protocols. Use only those articles and protocols mentioned in the website. \\
Website link: http://www.hri.org/docs/ECHR50.html\\
Steps to follow:\\
\{algo\}\\
\#\#\\
Court Statement:\\
```\{inp\}```\\
STRICTLY USE THE SPECIFIED STEPS FOR PREDICTION. DO NOT USE ANY OTHER ALGORITHM. If some articles or protocols are violated, mention them at the end with the relevant results from steps provided by me that led you to that conclusion.\\
Format of response: 1 \textbackslash n if any one of given article or protocol is violated and 0 \textbackslash n if none of them are violated. \\
The 1 (if any article or protocol is violated) or 0 (if no article or protocol is violated) should be on the first line followed by a newline character.\\
NOTE: THERE CAN BE MULTIPLE VIOLATIONS IN A SINGLE STATEMENT OR NONE. Return 0 if you are even slightly unsure if any right is violated as we dont want false convictions.\\
\hline
    \end{tabular}
    \caption{Robustness Check Prompt: Human Rights Violation (Llama3-70b). The algorithm is in Table \ref{tab:HR_algo_Llama}.}
    \label{tab:robustness_HR_Llama}
\end{table*}



\begin{table*}[]
    \centering

    \begin{tabular}{|p{16cm}|}
    \hline
       You are a highly skilled financial analyst with expertise in predicting stock price movements based on sentiment analysis. Your task is to determine whether the stock price of a given ticker will rise or fall based on the sentiment expressed in tweets related to that ticker.\\
Instructions:\\
            1. Analyze the provided tweets to assess the overall sentiment.\\
            2. If the overall sentiment is positive, predict that the stock price will rise (output 1).\\
            3. If the overall sentiment is negative, predict that the stock price will fall (output 0).\\
            Note 1: If you find any negative sentiment in the tweets, focus more on that.\\
            Most of the tweets are positive or neutral, so it is crucial to pay close attention to negative sentences. Negative sentiment carries more weight than positive sentiment in this analysis, so prioritize identifying and considering negative expressions but don't forget about positive ones.

\\

\hline
       
    \end{tabular}
    \caption{Task Prompt: Stock Prediction}
    \label{tab:stock_task_prompt}
\end{table*}

\begin{table*}[]
    \centering

    \begin{tabular}{|p{16cm}|}
    \hline
    **Robust Sentiment Analysis Algorithm**\\

**Input:** A set of tweets related to a given stock ticker.\\

**Output:** The overall sentiment of the tweets, represented by a numerical score.\\

**Steps:**\\

1. **Preprocess the tweets:** Remove irrelevant tweets and tokenize/lowercase the text of each tweet.
2. **Use a sentiment analysis tool to assign a sentiment score to each tweet:** I used the VADER (Valence Aware Dictionary and sEntiment Reasoner) sentiment analysis tool, which is specifically designed to analyze social media text. VADER assigns a compound sentiment score to each tweet, ranging from -1 (negative) to +1 (positive). VADER considers the lexical features of the text, as well as the presence of negations and intensifiers, to determine the sentiment.\\
3. **Calculate the overall sentiment:** Take a weighted average of the sentiment scores of all the tweets, where recent tweets are given more weight. I used a simple linear weighting scheme, where the weight of a tweet is inversely proportional to its age.
4. **Apply robustness measures:** To improve the robustness of the algorithm, I applied the following measures:\\
    * **Negation handling:** I identified and handled negations in the tweets to ensure that negative sentiment was correctly captured. For example, the tweet ``AAPL is not a good investment'' would be assigned a negative sentiment score, even though it contains the word ``not.''\\
    * **Contextual analysis:** I used VADER's ability to consider the context and tone of the tweets to improve the accuracy of the sentiment analysis. For example, the tweet ``AAPL is a great company, but their stock price is too high'' would be assigned a neutral sentiment score, even though it contains both positive and negative words.\\
5. **Return the overall sentiment score:** This score represents the overall sentiment of the tweets, which can be used to predict the stock price movement.\\

Algorithm End \\

**Example:**\\

Consider the following set of tweets related to AAPL:\\

* "\$AAPL - wall st. kicks off new year on lower note -> URL stock stocks stockaction" (Sentiment score: -0.4)
* ``rt AT\_USER our top 3 trade ideas for 2014 and two \$ 1,000 futures wins already in the bank \$ gld \$ uso \$ aapl - - URL'' (Sentiment score: 0.2)\\
* ''\$AAPL blackberry and singer alicia keys to part ways URL''(Sentiment score: -0.2)\\
* ``rt AT\_USER apple slipped, urban outfitters up on analyst call URL \$AAPL'' (Sentiment score: -0.3)
* ``\$AAPL - pre-market: apple downgraded on margin concerns ; stocks to start 2014 ... -> URL stock stocks stockaction'' (Sentiment score: -0.4)

Using the above algorithm, the overall sentiment score for these tweets would be:

```
Overall sentiment = (0.2 * 1 + (-0.4) * 0.8 + (-0.2) * 0.7 + (-0.3) * 0.6 + (-0.4) * 0.5) / (1 + 0.8 + 0.7 + 0.6 + 0.5) = -0.3
'''

This negative sentiment score indicates that the overall sentiment towards AAPL in these tweets is negative.
\\
\hline
    \end{tabular}
    \caption{ReQuest Algorithm: Stock Prediction}
    \label{tab:stock_algo}
\end{table*}

\begin{table*}[]
    \centering

    \begin{tabular}{|p{16cm}|}
    \hline
  You are a highly skilled financial analyst with expertise in predicting stock price movements based on sentiment analysis.\\
            Your task is to determine whether the stock price of a given ticker will rise or fall based on the sentiment expressed in
            tweets related to that ticker.\\
Instructions:\\
            1. Analyze the provided tweets to assess the overall sentiment.\\
            2. If the overall sentiment is positive, predict that the stock price will rise (output 1).\\
            3. If the overall sentiment is negative, predict that the stock price will fall (output 0).\\
            Note 1: If you find any negative sentiment in the tweets, focus more on that.\\
            Most of the tweets are positive or neutral, so it is crucial to pay close attention to negative sentences. Negative sentiment carries more weight than positive sentiment in this analysis, so prioritize identifying and considering negative expressions but dont forget about positive ones.\\ Note 2: To ensure consistency and reproducibility, adhere strictly to the following algorithm.Note that the provided example is intended only to demonstrate how the algorithm works; do not base your prediction solely on this example. \{algo\_prompt\}

\\
\hline
    \end{tabular}
    \caption{Robustness Check Prompt: Stock Prediction. The algorithm  is in Table \ref{tab:stock_algo}.}
    \label{tab:robustness_stock}
\end{table*}

\begin{table*}[]
    \centering

    \begin{tabular}{|p{16cm}|}
    \hline
       f``You are a skilled medical professional.\\
          You are an expert at predicting the level of depression severity of a given text.\\
          Your task is to choose a class from the list of depression severity levels of the text given : \{text\}.\\
          You can choose only one option from these classes : [\{classes\}],\\
          You must chose only one level from the list of classes given.\\
          Return only one class of depression severity level for the text given.''

   \\
\hline
       
    \end{tabular}
    \caption{Task Prompt: Reddit Suicide Watch (Gemini)}
    \label{tab:disease_task_prompt}
\end{table*}

\begin{table*}[]
    \centering

    \begin{tabular}{|p{16cm}|}
    \hline
       1. Collect a dataset of texts with known severity labels. This dataset can be obtained from a variety of sources, such as online forums, medical records, or research studies.\\
2. Preprocess the text data. This step involves removing punctuation, stop words, and other irrelevant characters from the text. The data is then converted to lowercase and tokenized.\\
3. Vectorize the text data. The tokenized text data is then vectorized using a technique such as bag-of-words or TF-IDF. This step converts the text data into a numerical format that can be processed by a machine learning algorithm.\\
4. Split the dataset into training and testing sets. The dataset is then split into two sets: a training set and a testing set. The training set is used to train the machine learning algorithm, and the testing set is used to evaluate the performance of the algorithm.\\
5. Train a machine learning algorithm. A machine learning algorithm, such as a support vector machine or a random forest, is then trained on the training set. The algorithm learns to identify the patterns in the text data that are associated with different severity labels.\\
6. Evaluate the performance of the machine learning algorithm. The performance of the machine learning algorithm is then evaluated on the testing set. The algorithm's performance is measured using a variety of metrics, such as accuracy, precision, and recall.\\
7. Deploy the machine learning algorithm. Once the machine learning algorithm has been trained and evaluated, it can be deployed to a production environment. The algorithm can be used to predict the severity of new texts as they are received.\\ 
\\

Here is an example of a simple algorithm that can be used to predict the severity of a text:\\ 
\\

def predict\_severity(text):\\
  """Predicts the severity of a text.\\

  Args:\\
    text: The text to predict the severity of.\\

  Returns:\\
    The predicted severity of the text.\\
  """

  \# Preprocess the text.\\
  text = preprocess\_text(text)\\

  \# Vectorize the text.\\
  vector = vectorize\_text(text)\\

  \# Load the machine learning model.\\
  model = load\_model("severity\_prediction\_model.pkl")\\

  \# Predict the severity of the text.\\
  severity = model.predict(vector)\\

  \# Return the predicted severity.\\
  return severity\\
\\
\hline
    \end{tabular}
    \caption{ReQuest Algorithm: Reddit Suicide Watch (Gemini)}
    \label{tab:disease_algo}
\end{table*}

\clearpage 

\begin{table*}[t]
    \centering

    \begin{tabular}{|p{16cm}|}
    \hline
      f``You are an expert at following algorithms. You strictly adhere to the algorithm provided.\\
          Your task is to predict the class of depression severity from the text given using this algorithm step by step.\\
          Algorithm: \{Reproducible\_algorithm\}.\\
          You have to predict a class from this list = \{classes\}.\\
          You must choose a class of severity for this text : {text}.\\
          You must choose only one class from the list of classes given.\\
          Only return the class of depression severity of the text from the list of classes.\\
          No need to provide any explanation.''
\\
\hline
    \end{tabular}
    \caption{Robustness Check Prompt (Gemini): Reddit Suicide Watch. The algorithm  is in Table \ref{tab:disease_algo}.}
    \label{tab:robusteness_disease}
\end{table*}

\begin{table*}[]
    \centering

    \begin{tabular}{|p{16cm}|}
    \hline
        f``You are a skilled medical professional.\\
          You are an expert at predicting the class of a given text.\\
          Your task is to choose a class from the list of nature of types of the text given : \{text\}.\\
          You can choose only one option from these classes : [\{classes\}],\\
          You must chose only one type from the list of classes given.\\
          Return only one class type for the text given.''
                \\
\hline
       
    \end{tabular}
    \caption{Task Prompt: Depression Severity Detection (Gemini)}
    \label{tab:medicine_task_prompt}
\end{table*}

\onecolumn
\begin{longtable}{|p{16cm}|}
    \hline
    Step 1: Data Preprocessing \\
    1. Tokenization: The input text is tokenized to break it into individual words or tokens. \\
    2. Stop word removal: Common words that do not contribute to meaning, such as "the," "and," and "is," are removed. \\
    3. Stemming or Lemmatization: Words are reduced to their root forms to capture their core meaning and avoid overfitting on different word forms. \\
    Step 2: Feature Extraction \\
    1. Bag-of-Words (BoW) or TF-IDF: The frequency of each token in the text is calculated to create a feature vector that represents the text. TF-IDF (Term Frequency-Inverse Document Frequency) weighting can be used to emphasize the importance of tokens that are frequent in the given text but rare across a corpus of texts. \\
    2. N-grams: Sequences of adjacent tokens (e.g., bigrams or trigrams) can be extracted to capture phrases or expressions that convey specific sentiments or topics. \\
    3. Sentiment Analysis: The overall sentiment of the text can be analyzed using tools like VADER (Valence Aware Dictionary and Sentiment Reasoner) or TextBlob, which assign valence scores to tokens to determine the emotional tone of the text. \\
    Step 3: Model Training \\
    1. Classification Algorithm: A supervised machine learning algorithm is chosen, such as Logistic Regression, Support Vector Machines (SVM), or Naive Bayes. \\
    2. Training Data: The algorithm is trained on a labeled dataset of text and class labels (e.g., SuicideWatch, Depression, Anxiety). \\
    3. Model Parameters: The algorithm's hyperparameters (e.g., regularization strength, kernel function) are tuned to optimize performance on a validation set. \\
    Step 4: Model Evaluation \\
    1. Evaluation Metrics: Metrics such as accuracy, precision, recall, and F1-score are used to assess the performance of the model on a test set. Model Selection: The best performing model is selected based on the evaluation results. \\
    Step 5: Classification of New Text \\
    1. Preprocessing: New text is processed using the same preprocessing steps as the training data. \\
    2. Feature Extraction: Features are extracted from the new text using the same techniques used for training. \\
    3. Prediction: The trained model is used to predict the class label of the new text. \\
    Example Algorithm in Python using Scikit-Learn \\
    from sklearn.feature\_extraction.text import CountVectorizer, TfidfVectorizer \\
    from sklearn.linear\_model import LogisticRegression \\
    from sklearn.model\_selection import train\_test\_split, cross\_val\_score \\
    \# Import the training data \\
    train\_data = pd.read\_csv('training\_data.csv') \\
    \# Preprocess the text \\
    train\_data['text'] = train\_data['text'].str.lower() \\
    \# Create a bag-of-words feature extractor \\
    vectorizer = CountVectorizer(stop\_words='english') \\
    \# Extract features from the training data \\
    X\_train = vectorizer.fit\_transform(train\_data['text']) \\
    \# Create a Logistic Regression classifier \\
    classifier = LogisticRegression() \\
    \# Train the classifier \\
    classifier.fit(X\_train, train\_data['class\_label']) \\
    \# Evaluate the classifier \\
    scores = cross\_val\_score(classifier, X\_train, train\_data['class\_label'], cv=5) \\
    print('Cross-validation score:', np.mean(scores)) \\
    \# Predict the class label of a new text \\
    new\_text = 'I feel hopeless and like I want to give up on life.' \\
    new\_features = vectorizer.transform([new\_text]) \\
    prediction = classifier.predict(new\_features) \\
    print('Predicted class:', prediction) \\
    \hline
    \caption{ReQuest Algorithm:  Depression Severity Detection (Gemini)}
    \label{tab:medicine_algo}
\end{longtable}
\twocolumn  

\begin{table*}[htbp]
    \centering

    \begin{tabular}{|p{16cm}|}
    \hline
   prompt = f``You are an expert at following algorithms. You strictly adhere to the algorithm provided.\\
          Your task is to predict the class of nature of the given text using this algorithm step by step.\\
          Algorithm: \{Reproducible\_algorithm\}.\\
          You have to predict a class from this list = \{classes\}.\\
          You must choose a class type for this text : \{text\}.\\
          You must choose only one class from the list of classes given.\\
          Only return the class of nature of given text from the list of classes.\\
          No need to provide any explanation.\\

\hline
    \end{tabular}
    \caption{Robustness Check Prompt (Gemini): Depression Severity Detection. The algorithm  is in Table \ref{tab:medicine_algo}.}
    \label{tab:robustness_medicine}
\end{table*}

\begin{figure*}[] 
  \centering    
\includegraphics[width=0.75\textwidth, height=8cm]{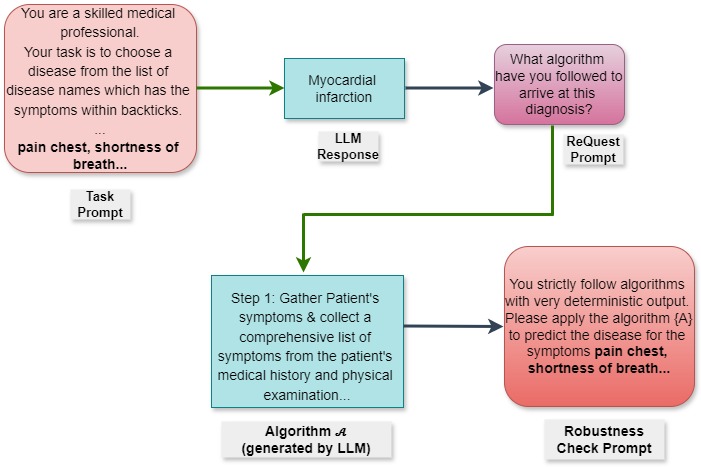} 
  \caption{An example of the novel prompt regime for the health domain.} 
  \label{fig:request_eg}
\end{figure*}

\begin{table*}[]
    \centering

    \small
    \begin{tabular}{|c|p{2cm}|p{5.5cm}|c|}
    \hline
       {\bf Dataset}  & {\bf Task} & {\bf Example Input (excerpt)} & {\bf Example Output}\\
       \hline
        \cite{vats2023llms} & Statute Prediction &  ..lodged an FIR alleging that the husband and the mother-in-law of the deceased, after the marriage, had been constantly {\color{red}asking for dowry of Rs. 2 lakh} from the father of the deceased.. & Penalty for demanding dowry\\
        \hline
     \cite{chalkidis-etal-2019-neural} & Human Rights Violation Prediction & ...understood that police officers had {\color{red}handcuffed the applicant and hung him and afterwards hit his head against the wall}. According to V., police officers had handcuffed the applicant and hung him, and then either he himself... & Human rights violated\\
     \hline
    \end{tabular}
    \caption{Examples from the legal datasets}
    \label{tab:legal_egs}
\end{table*}

\begin{table*}[]
    \centering

    \begin{tabular}{|c|p{2cm}|p{5.5cm}|c|}
    \hline
    \multicolumn{4}{|c|}{Health Tasks}\\
    \hline
       {\bf Dataset}  & {\bf Task} & {\bf Example Input (excerpt)} & {\bf Example Output}\\
       \hline
       \cite{naseem2022early} & Suicidal \hspace{3mm} tendencies Classification &  ..hated the fact that I have these mean voices in my head telling me that {\color{red}no one truly cares about me}. I have {\color{red}hated the fact that I was so insecure about my weight}.. & SuicideWatch\\
        \hline
     \cite{ji2021suicidal}. & Depression Severity Detection & ..then {\color{red}hit me with the newspaper} and it shocked me that she would do this, she knows I don\'t like {\color{red}play hitting, smacking, striking, hitting or violence}.. & Moderate\\
     \hline
      \multicolumn{4}{|c|}{Finance Tasks}\\
    \hline
       {\bf Dataset}  & {\bf Task} & {\bf Example Input (excerpt)} & {\bf Example Output}\\
       \hline
       \cite{xu-cohen-2018-stock} & Stock Price Prediction &  ..here's how apple could be making a huge push into healthcare..	 & Price increase (Apple)\\
        \hline
     
     \hline
    \end{tabular}
    \caption{Examples from the health and finance datasets}
    \label{tab:health_finance_egs}
\end{table*}

\begin{table*}[h]
\centering
\makebox[\textwidth][c]{ 
\rowcolors{1}{tablebg}{white}
\tiny
\begin{tabular}{|c|c|c|c|c|c|c|c|c|c|}
\hline
\rowcolor{lightred}
\multicolumn{10}{|c|}{Google (GOOG)} \\ 
\hline
\multicolumn{5}{|c|}{Intra-LLM} & \multicolumn{5}{|c|}{Inter-LLM} \\ 
\hline
\multicolumn{3}{|c|}{Macro F1} &  Percentage & Ratio & Macro F1 &  Percentage & Ratio &  Percentage & Ratio\\ 
\hline
{\bf baseline (BERTweet)} & {\bf Gemini} & {\bf ReQuest algo} &  {\bf PerRR\_GP\_G} & {\bf PreRR\_GP\_G} & {\bf Gemini $\rightarrow$ Llama} & {\bf PerRR\_GP\_L} & {\bf PreRR\_GP\_L} & {\bf PerRR\_GA\_L} & {\bf PreRR\_GA\_L}\\
\hline
                   0.55            &  0.45          &  0.46        & 97.82 {\color{green}$\uparrow$} & 0.5464 & 0.48 & 93.75 &  0.6473 & 95.83 {\color{green}$\uparrow$} & 0.6843\\
\hline                   
 {\bf baseline (BERTweet)} & {\bf Llama} & {\bf ReQuest algo} &  {\bf PerRR\_LP\_L} & {\bf PreRR\_LP\_L} & {\bf Llama $\rightarrow$ Gemini} & {\bf PerRR\_LP\_G} & {\bf PreRR\_LP\_G} & {\bf PerRR\_LA\_G} & {\bf PreRR\_LA\_G}\\
\hline
 0.55             & 0.37       &  0.35         & 94.59 {\color{red}$\downarrow$} & 0.9746 & 0.52 & 71.15 &  0.6052 & 67.30 {\color{green}$\uparrow$} & 0.5502\\
\hline
\rowcolor{lightred}
\multicolumn{10}{|c|}{Amazon (AMZN)} \\ 
\hline
\multicolumn{5}{|c|}{Intra-LLM} & \multicolumn{5}{|c|}{Inter-LLM} \\ 
\hline
\multicolumn{3}{|c|}{Macro F1} &  Percentage & Ratio & Macro F1 &  Percentage & Ratio &  Percentage & Ratio\\ 
\hline
{\bf Baseline (BERTweet)} & {\bf Gemini} & {\bf ReQuest algo} &  {\bf PerRR\_GP\_G} & {\bf PreRR\_GP\_G} & {\bf Gemini $\rightarrow$ Llama} & {\bf PerRR\_GP\_L} & {\bf PreRR\_GP\_L} & {\bf PerRR\_GA\_L} & {\bf PreRR\_GA\_L}\\
\hline
0.53 & 0.47 & 0.49 & 95.91 {\color{green}$\uparrow$} & 0.4915 & 0.39 & 82.97 & 0.3389 & 79.59{\color{red}$\downarrow$} & 0.6207\\
\hline                   
{\bf Baseline (BERTweet)} & {\bf Llama} & {\bf ReQuest algo} &  {\bf PerRR\_LP\_L} & {\bf PreRR\_LP\_L} & {\bf Llama $\rightarrow$ Gemini} & {\bf PerRR\_LP\_G} & {\bf PreRR\_LP\_G} & {\bf PerRR\_LA\_G} & {\bf PreRR\_LA\_G}\\
\hline
0.53 & 0.43 & 0.37 & 86.04 {\color{red}$\downarrow$} & 0.5021 & 0.49 & 87.75 & 0.4131 & 75.51{\color{red}$\downarrow$} & 0.3512\\
\hline
\end{tabular}
} 
\caption{
Reproducibility: Finance - Google and Amazon datasets}
\label{tab:financeadditional}
\end{table*}


\begin{figure*}[] 
  \centering    
\includegraphics[width=0.75\textwidth, height=8cm]{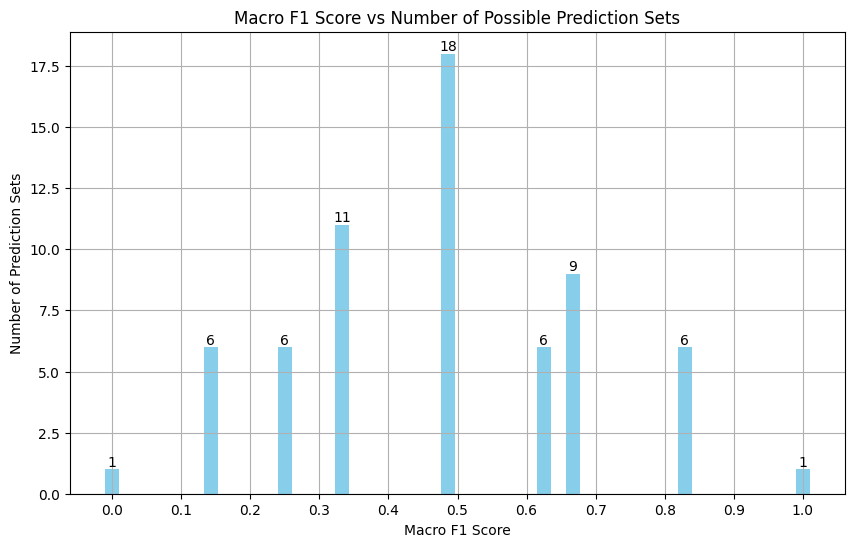} 
  \caption{Distribution showing no of possible unique prediction sets vs Macro-F1 for a balanced binary dataset of size 6. The highest ambiguity in Macro-F1 score is towards the centre.} 
  \label{fig:balanced_binary}
\end{figure*}

\begin{figure*}[] 
  \centering    
\includegraphics[width=0.75\textwidth, height=8cm]{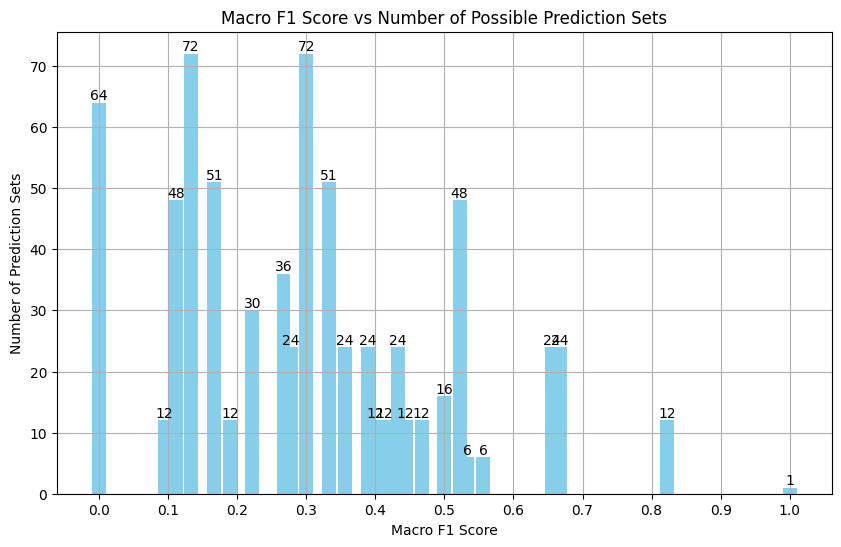} 
  \caption{Distribution showing no of possible unique prediction sets vs Macro-F1 for a balanced 3-class multi-class dataset of size 6. Most of the ambiguity in Macro-F1 score is now shifted left as compared to a relatively simpler binary classification problem.} 
  \label{fig:multi_class}
\end{figure*}

\end{document}